\documentclass[prapplied,12pt,superscriptaddress,nofootinbib,notitlepage,longbibliography]{revtex4-1}

\usepackage{soul}
\usepackage{amsthm}
\usepackage{amsmath}
\usepackage{amssymb}
\usepackage{graphicx}
\usepackage{epstopdf}
\usepackage{subfig}
\usepackage{url}
\usepackage{float}
\usepackage{bm}
\usepackage{ifthen}
\usepackage[usenames,dvipsnames]{color}
\usepackage{mathrsfs}
\usepackage[colorlinks=false,citecolor=blue,urlcolor=black]{hyperref}
\usepackage{float}

\usepackage{placeins}
\newcommand{\sket}[1]{{\ensuremath{\lvert#1\rangle}}}
\newcommand{\lket}[1]{{\ensuremath{\left\lvert#1\right\rangle}}}
\newcommand{\ket}[1]{\if@display\lket{#1}\else\sket{#1}\fi}

\newcommand{\sbra}[1]{{\ensuremath{\langle#1\rvert}}}
\newcommand{\lbra}[1]{{\ensuremath{\left\langle#1\right\rvert}}}
\newcommand{\bra}[1]{\if@display\lbra{#1}\else\sbra{#1}\fi}

\begin{document}
\title{Time-bin and Polarization Superdense Teleportation for Space Applications}
\author{Joseph C. Chapman}
\email{jchapmn2@illinois.edu}
\affiliation{Illinois Quantum Information Science and Technology Center, University of Illinois at Urbana-Champaign, Urbana, IL 61801}
\affiliation{Dept. of Physics, University of Illinois at Urbana-Champaign, Urbana, IL 61801}
\author{Trent M. Graham}
\affiliation{Dept. of Physics, University of Illinois at Urbana-Champaign, Urbana, IL 61801}
\affiliation{Dept. of Physics, University of Wisconsin - Madison, Madison, WI 53706}
\author{Christopher K. Zeitler}
\affiliation{Illinois Quantum Information Science and Technology Center, University of Illinois at Urbana-Champaign, Urbana, IL 61801}
\affiliation{Dept. of Physics, University of Illinois at Urbana-Champaign, Urbana, IL 61801}
\author{Herbert J. Bernstein}
\affiliation{Institute for Science \& Interdisciplinary Studies \& School of Natural Sciences, Hampshire College, Amherst, MA 01002}
\author{Paul G. Kwiat}
\affiliation{Illinois Quantum Information Science and Technology Center, University of Illinois at Urbana-Champaign, Urbana, IL 61801}
\affiliation{Dept. of Physics, University of Illinois at Urbana-Champaign, Urbana, IL 61801}
%\date{}

%\abstract{To build a global quantum communication network, low-transmission, fiber-based communication channels can be supplemented by using a free-space channel between a satellite and a ground station on Earth. We have constructed a system that generates hyperentangled photonic ``ququarts'' and measures them to execute multiple quantum communication protocols of interest. We have successfully executed and characterized superdense teleportation---our measurements show an average fidelity of $0.94\pm0.02$, with a phase resolution of $\sim7^{\circ}$, allowing reliable transmission of $>10^5$ distinguishable quantum states. Additionally, we investigated the fidelity and phase uncertainty dependence on photon count. We found that a typical pass of of the low-earth orbit satellite should collect. Finally, we have demonstrated the ability to compensate for the Doppler shift, which would otherwise prevent sending time-bin encoded states from a rapidly moving satellite, thus allowing the low-error execution of phase-sensitive protocols during an orbital pass. }
\begin{abstract}
%{To build a global quantum communication network, low-transmission, fiber-based channels can be supplemented by using a free-space channel between a satellite and Earth. To this end, we constructed a system that generates hyperentangled photonic ``ququarts'' and used them for superdense teleportation. Our measurements show an average fidelity of $0.94\pm0.02$, with a phase resolution of $\sim7^{\circ}$, allowing reliable transmission of $>10^5$ distinguishable quantum states. Finally, we demonstrate that the protocol can be reliably executed due to a sufficient estimated photon-flux and Doppler shift compensation.}
To build a global quantum communication network, low-transmission, fiber-based communication channels can be supplemented by using a free-space channel between a satellite and a ground station on Earth. We have constructed a system that generates hyperentangled photonic ``ququarts'' and measures them to execute multiple quantum communication protocols of interest. We have successfully executed and characterized superdense teleportation, a modified remote-state preparation protocol that transfers more quantum information than standard teleportation, for the same classical information cost, and moreover, is in principle deterministic. Our measurements show an average fidelity of $0.94\pm0.02$, with a phase resolution of $\sim7^{\circ}$, allowing reliable transmission of $>10^5$ distinguishable quantum states. Additionally, we have demonstrated the ability to compensate for the Doppler shift, which would otherwise prevent sending time-bin encoded states from a rapidly moving satellite, thus allowing the low-error execution of phase-sensitive protocols during an orbital pass. Finally, we show that the estimated number of received coincidence counts in a realistic implementation is sufficient to enable faithful reconstruction of the received state in a single pass. 
\end{abstract}
\maketitle
%\tableofcontents

\section{Introduction}
A global quantum network would have myriad uses. For example, it could improve the collective computational power of quantum computers by allowing them to communicate \cite{DistQC}, enable arbitrarily long-distance secure communication using quantum cryptography \cite{SatQN}, and might even facilitate planet-scale distributed quantum sensors, e.g., for clock synchronization \cite{giovannetti2001quantum}, and super-resolution telescopy \cite{QVLBI,Lukin1,Lukin2}. Currently, the distance between nodes in a potential quantum network is limited by the absorption loss in fiber-optic cables or the effects of turbulence for free-space terrestrial channels. If a channel between a satellite and Earth were used as part of the network, the distances between nodes could be greatly increased because that channel is less affected by the above limiting factors \cite{simon2017towards}. The utility of a free-space satellite-Earth channel has been recognized by many research groups around the world  \cite{optphotnews}. For example, the Chinese Micius satellite was used to demonstrate long-distance photonic-entanglement distribution \cite{entanglement2017sat}, one version of quantum key distribution (QKD)  \cite{qkd2017sat,liao2018satellite}, and a preliminary test of quantum teleportation  \cite{tel2017ground}. Additionally, there is significant work ongoing in Singapore  \cite{grieve2018spooqysats}, Italy  \cite{vallone2015experimental}, Canada  \cite{pugh2017airborne}, and Austria  \cite{steinlechner2017distribution,liao2018satellite}.

To further the development of quantum communication applications in space, we have created a system that can execute multiple quantum communication protocols, including  high-dimensional entanglement-based quantum key distribution \cite{chapman2019hyperentangled}, superdense teleportation (SDT)  \cite{SDTHerb}, {and high-dimensional Bell inequality and quantum steering tests \cite{zeitlerbell}. With additional modifications, the system could be suitable for operation on a satellite. In this work, we characterize our source of hyperentangled photons and the performance of SDT in our system over its whole message space, and demonstrate through lab test and calculation the ability to robustly execute SDT during a single orbital pass of a low-earth orbit satellite, e.g., the International Space Station (ISS). This work significantly exceeds previous SDT work} \cite{SDTOAM} {by the authors, due to incorporation of robust quantum degrees of freedom/elimination of encoding in non-robust degrees of freedom, compensation for the Doppler shift (which otherwise renders the protocol useless), and a more thorough characterization of the SDT system in general; we have also incorporated a comparison between the standard Maximum Likelihood analysis and Bayesian analysis methods that are superior to standard quantum state tomography techniques for low-rate communications. Finally, for the first time we have undertaken a thorough analysis to project rates and performance in a low-earth-orbit demonstration of this protocol.}

\section{Superdense Teleportation Protocol}
\begin{figure*}
\centerline{\includegraphics[scale=0.6]{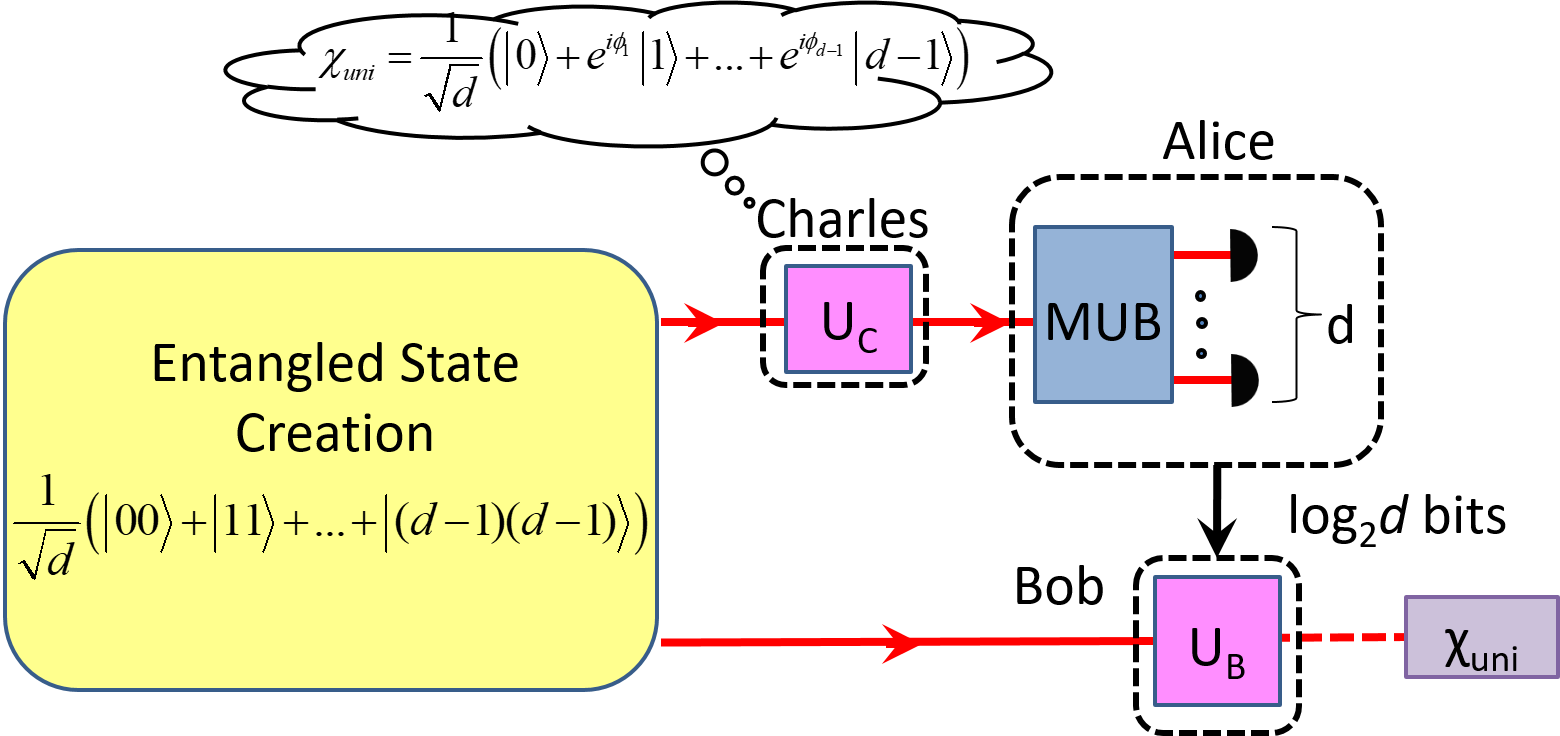}}\caption{{Superdense teleportation (SDT): Charles desires to send Bob a d-dimensional equimodular state. To do so via SDT, Charles and Bob start by sharing a d-dimensional maximally entangled state; Charles makes local unitary operations to set the desired phases of the equimodular state. Alice then makes a measurement in a mutually-unbiased basis to the basis Charles applied his phases in. Her measurement outcome is transmitted to Bob, who then makes the necessary unitary transformation to deterministically convert his half of the entangled pair into the precise state Charles wanted to send.}}\label{simplesdt}
\end{figure*}

SDT, as shown in Fig. \ref{simplesdt}, {is a three-party protocol involving Alice, Bob, and Charles. Charles wants to send Bob an ``equimodular'' quantum state---a subset of the states in the available Hilbert space, in which all terms have the same magnitude:}
\begin{equation}
\ket{\Psi_C}= \frac{1}{\sqrt{d}}(\ket{0}+e^{i\phi_1}\ket{1}+e^{i\phi_2}\ket{2}+...e^{i\phi_{d-1}}\ket{(d-1)})
\end{equation} for any values of $\phi_1$, $\phi_2$, $... \phi_{d-1}$ $\in$  $[0,2\pi)$. To begin the protocol, Bob and Charles share a d-dimensional maximally entangled state:
\begin{equation}
\ket{\Psi_{BC}}= \frac{1}{\sqrt{d}}(\ket{00}+\ket{11}+\ket{22}+...\ket{(d-1)(d-1)})\text{,} \label{psibc}
\end{equation}
onto which Charles locally encodes his desired phases:
\begin{equation}
\ket{\Psi_{BC}}= \frac{1}{\sqrt{d}}(\ket{00}+e^{i\phi_1}\ket{11}+e^{i\phi_2}\ket{22}+...e^{i\phi_{d-1}}\ket{(d-1)(d-1)})\text{.}\label{psibcphi}
\end{equation}
 Next, Alice measures Charles' photon in a mutually-unbiased basis from the one in which Charles applied the phases, e.g., the basis
\begin{align}
\ket{A_1}&\equiv\frac{1}{2}(+\ket{0}+\ket{1}+\ket{2}-\ket{3})\\
\ket{A_2}&\equiv\frac{1}{2}(+\ket{0}+\ket{1}-\ket{2}+\ket{3})\\
\ket{A_3}&\equiv\frac{1}{2}(+\ket{0}-\ket{1}+\ket{2}+\ket{3})\\
\ket{A_4}&\equiv\frac{1}{2}(-\ket{0}+\ket{1}+\ket{2}+\ket{3})\text{,}
\label{Alicestates}
\end{align}
 where we now restrict our discussion to d=4, relevant for our experimental implementation. States $\ket{A_1}$ to $\ket{A_4}$ are those projected onto by Alice's 4 detectors.  Before Alice's measurement, the full state of the system is given by 
\begin{align}
\ket{\Psi_{AB}}=\frac{1}{2}\Big(\frac{1}{2}\ket{A_1}&\otimes(+\ket{0}+e^{i\phi_1}\ket{1}+e^{i\phi_2}\ket{2}-e^{i\phi_3}\ket{3})\notag\\
+\frac{1}{2}\ket{A_2}&\otimes(+\ket{0}+e^{i\phi_1}\ket{1}-e^{i\phi_2}\ket{2}+e^{i\phi_3}\ket{3})\notag\\
+\frac{1}{2}\ket{A_3}&\otimes(+\ket{0}-e^{i\phi_1}\ket{1}+e^{i\phi_2}\ket{2}+e^{i\phi_3}\ket{3})\notag\\
+\frac{1}{2}\ket{A_4}&\otimes(-\ket{0}+e^{i\phi_1}\ket{1}+e^{i\phi_2}\ket{2}+e^{i\phi_3}\ket{3})\Big)\text{.}\label{psiab}
\end{align}
Upon measurement, Alice sends her result to Bob (using 2 classical bits of information), who then applies the correct unitary transformation (a $\pi$ phase shift on one of the four terms) so that  \cite{Trentthesis}
\begin{equation}
\ket{\Psi_B}=\ket{\Psi_C}=\frac{1}{2}(\ket{0}+e^{i\phi_1}\ket{1}+e^{i\phi_2}\ket{2}+e^{i\phi_3}\ket{3})\text{.}
\end{equation}

Although the SDT state space is restricted to equimodular states, and is therefore not suitable for general quantum computation, these states are sufficient to enable blind quantum computing, a client-server cluster quantum computing model that ensures privacy of the inputs, the outputs, and the computation being performed \cite{BQC}. Moreover, the SDT protocol is deterministically successful, in contrast to quantum teleportation and probabilistic remote state preparation, which both only succeed at most half of the time using linear optics \cite{SDTOAM}. In addition, it also uses fewer classical communication resources than quantum teleportation and deterministic remote state preparation \cite{SDTOAM}; for example, whereas standard teleportation requires Alice to send 2 classical bits to teleport a single qubit (described by two continuous variables, e.g., $\ket{\psi}=\cos{\theta}\ket{0}+\sin{\theta}e^{i\phi}\ket{1}$), SDT transmits three continuous variables for the same two classical bits. Higher-dimensional quantum teleportation has been performed before but with lower fidelity and only probabilistic success  \cite{wang2015quantum,luo2019quantum}. Furthermore, Alice's measurements for SDT are substantially less resource intensive than those needed, e.g., for remote state preparation \cite{SDTOAM}; this is an important consideration for a satellite-based protocol.
\section{SDT Protocol Execution}
\begin{figure*}
\centerline{\includegraphics[scale=0.6]{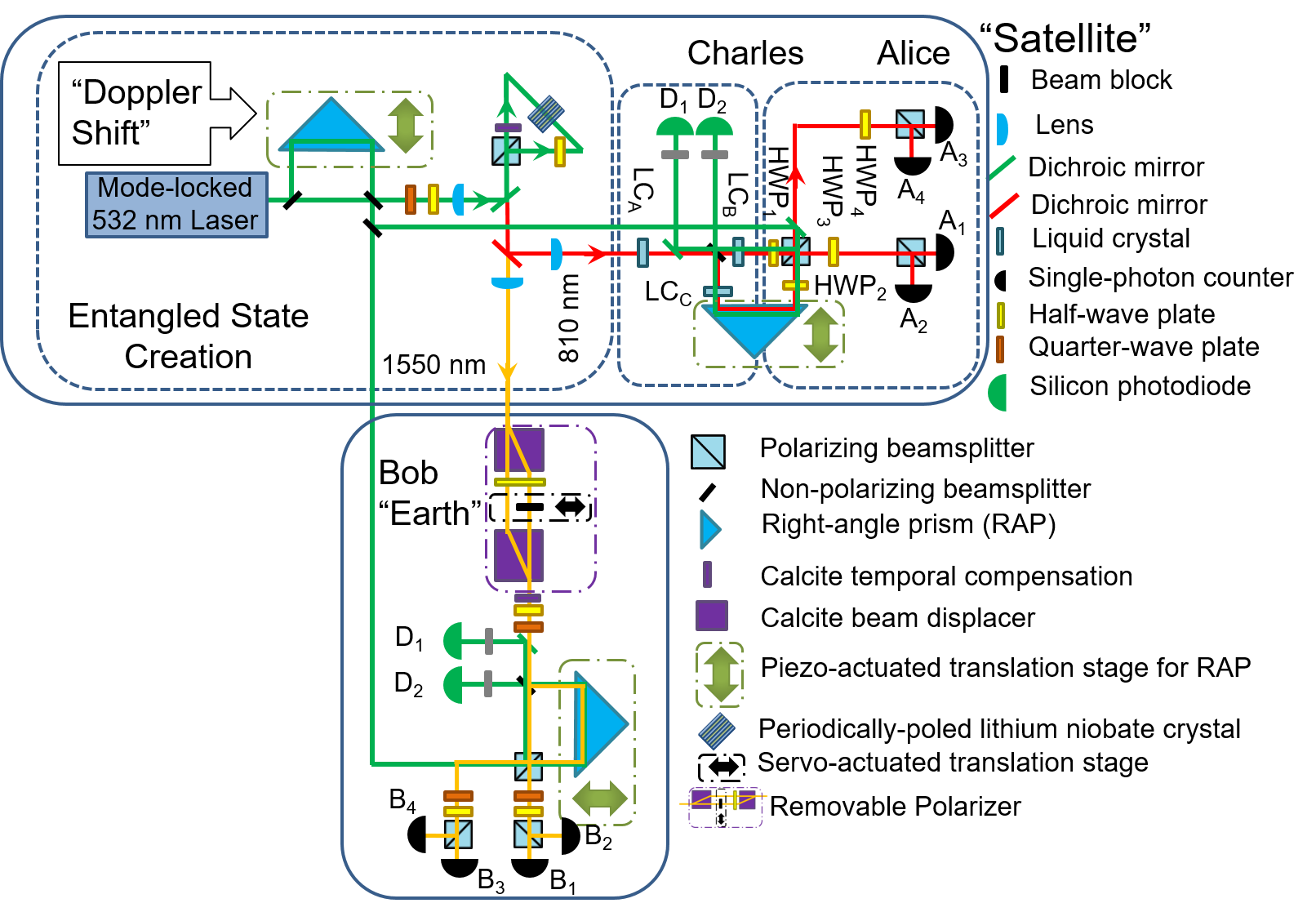}}\caption{Superdense teleportation optical setup: Photonic time-bin and polarization ququarts are generated via spontaneous parametric downconversion in periodically-poled lithium niobate (See Appendix \ref{stategen}). Green lines are the 532-nm pump (and stabilization) beam; red and yellow are  the signal (810 nm) and idler (1550 nm) photons, respectively. Each half of the ququart is manipulated and measured independently via Alice/Charles' and Bob's sections, which include full polarization analysis, complete phase manipulation, and an unbalanced interferometer to allow measurements of superpositions of time bins.}\label{sdtsetup}
\end{figure*}

Superdense teleportation has been executed previously using photons hyperentangled in their polarization and orbital angular momentum (OAM) \cite{SDTOAM}. For our intended goal of transmitting quantum information over a channel from space to earth, time bins are a much better choice than OAM modes, as the latter are corrupted by atmospheric turbulence and require larger apertures to faithfully detect  \cite{torres2012optical}. Using nondegenerate spontaneous parametric downconversion, our source produces time-bin and polarization entangled photons (see Appendix \ref{stategen}) in the 16-dimensional equimodular state 
\begin{equation}
\ket{\Psi_{BC}}= \frac{1}{2}(\ket{(Ht_1)_{810}(Ht_1)_{1550}}+\ket{(Vt_1)_{810}(Vt_1)_{1550}}+\ket{(Ht_2)_{810}(Ht_2)_{1550}}+\ket{(Vt_2)_{810}(Vt_2)_{1550}})\text{.} \label{psibcht}
\end{equation}
Here H and V refer to horizontal and vertical polarization, $t_1$ and $t_2$ refer to two time bins, and 810 and 1550 (nm) are the photon wavelengths. As shown in Fig. \ref{sdtsetup}, the 810-nm (1550-nm) photon is distributed to Charles (Bob), who applies phases $\phi_{LC_A}$, $\phi_{LC_B}$, and $\phi_{LC_C}$ on the state by actuating 3 different liquid crystals, $LC_A$, $LC_B$, and $LC_C$, to allow arbitrary phase selection over the range $[0,2\pi)$. Alice's projective measurement in a mutually unbiased basis is carried out by the polarizing beamsplitter (PBS) of her interferometer, preceded by $HWP_1$ and $HWP_2$ in the interferometer arms, which effectively place the PBS into the diagonal/anti-diagonal basis. Similarly, $HWP_3$ and $HWP_4$ are oriented at 22.5$^{\circ}$ with respect to horizontal, so the detectors project onto a superposition of time bins. Instead of having Bob complete the protocol by making the necessary unitary transformation, Bob measures 4 tomographies, conditioned on which of Alice's detectors fires. This allows us to tomographically reconstruct all 4 of the different states sent to Bob and apply the unitary transformation during the analysis after state reconstruction. 
\section{Results}
\begin{figure}
\centerline{\includegraphics[scale=1]{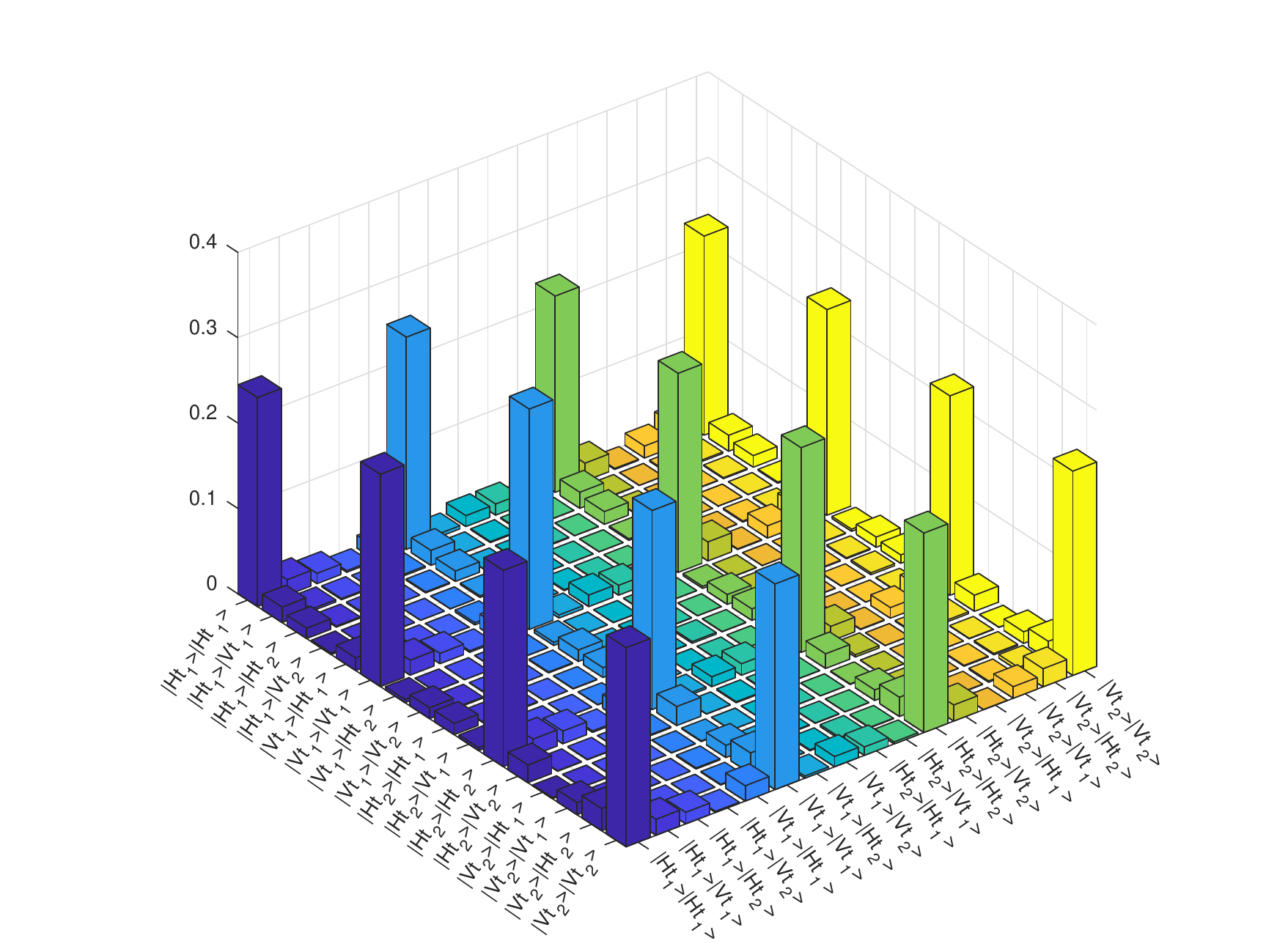}}\caption{Hyperentangled state density matrix: Absolute value of the reconstructed hyperentangled state density matrix, with a fidelity of $0.955\pm0.004$ with the desired state, and a purity of $0.932\pm0.007$.}\label{fulltomodenmat}
\end{figure}

To verify the quality of our source, we performed a full tomography of the joint state of the system by performing 1296 measurements (6 measurements for each qubit in the ququart). This involved adding extra half- and quarter-wave plates and a removable polarizer into Alice/Charles' side so a complete tomography could be made on each photon of the pair; see Appendix \ref{fulltomosec} for details. The purity of the reconstructed density matrix (Fig. \ref{fulltomodenmat}) is $P\equiv\text{Tr} \rho^2=0.932\pm0.007$; the fidelity of the absolute value of the reconstructed density matrix, $|\rho_m|$, with Eqn. \ref{psibcht} is 
\begin{equation}
F=\Bigg(\text{Tr}\sqrt{\rho_{BC}^{1/2}\text{ }|\rho_m|\text{ }\rho_{BC}^{1/2}}\Bigg)^2=0.955\pm0.004\text{.}
\end{equation} 
For each calculation, the error bar was produced from a Monte Carlo analysis, assuming Poissonian counting statistics and  using 100 samples, with mean count values on the order of the number of detected events for each of the tomography measurements (See Appendix \ref{tomo}, Table \ref{countfile}) .

\subsection{Phase Space Characterization}
\begin{figure}
\centerline{\includegraphics[height=4.7in,width=9in]{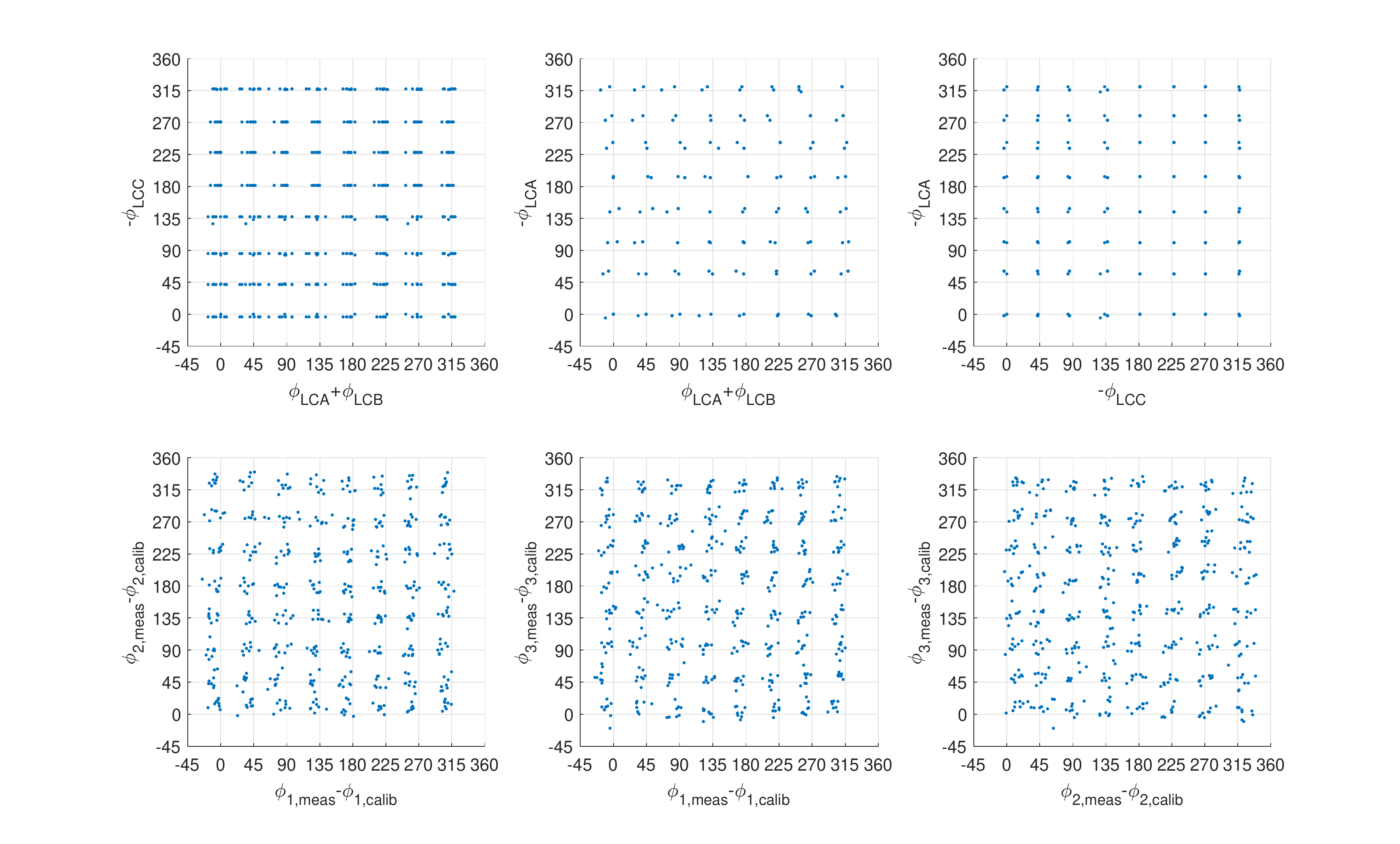}}\caption{Projections onto $45^{\circ}$ grid: All axes are in units of degrees. (top row) Target projections based on liquid crystal phase calibration. (bottom row) Measured phase projections after subtracting calibrated phase offset.}\label{45expmeas}
\end{figure}

To characterize the performance of the SDT protocol over the complete space of possible states (any value of $\phi_1$, $\phi_2$, and $\phi_{3}$ $\in$  $[0,2\pi)$), we measured every combination of $\phi_1$, $\phi_2$, and $\phi_{3}$ at roughly $45^{\circ}$ intervals between $[0,2\pi)$. These 512 states are represented in phase space in Fig. \ref{45expmeas} for trials where Alice obtained a click at detector $A_1$ (which is representative of events detected by the other three detectors). More precisely, the plots show the difference between the phases of the reconstructed state ($\phi_{1,meas}$, $\phi_{2,meas}$, and $\phi_{3,meas}$) and the calibration phases ($\phi_{1,calib}$, $\phi_{2,calib}$, and $\phi_{3,calib}$) from a calibration tomography taken every 4 tomographies. The phases between the polarizations are relatively stable and do not need to be measured except when the alignment changes, but the time-bin phase is more susceptible to slight phase drift. We believe the increased variation in $\phi_{LC_A}+\phi_{LC_B}$ (top left graph of Fig. \ref{45expmeas}) is due to the compounded variation in $\phi_{LC_A}$ and $\phi_{LC_B}$ as $\phi_{LC_A}$, the phase which changes within each grouping in the projections of Fig. \ref{45expmeas}, is increased. The average fidelity over the entire grid and all of Alice's detectors is
\begin{equation} F=\Bigg(\text{Tr}\sqrt{\rho_{tar}^{1/2}\rho_{meas}\rho_{tar}^{1/2}}\Bigg)^2=0.94\pm0.02\text{,}
\end{equation}
where $\rho_{tar}=\ket{\Psi_{tar}}\bra{\Psi_{tar}}$ and \begin{equation}
\ket{\Psi_{tar}}=\frac{1}{2}(\ket{0}+e^{i(\phi_{1,tar})}\ket{1}+e^{i(\phi_{2,tar})}\ket{2}+e^{i(\phi_{3,tar})}\ket{3})\text{.}
\end{equation}
Here
\begin{align}
\phi_{1,tar}&\equiv\phi_{1,calib}+\phi_{LC_A}+\phi_{LC_B}\\
\phi_{2,tar}&\equiv\phi_{2,calib}-\phi_{LC_C}\\
\phi_{3,tar}&\equiv\phi_{3,calib}+\phi_{LC_A}\text{.}
\end{align}
From the grids in Fig. \ref{45expmeas}, we calculate the standard deviation of $\Delta\phi_i\equiv\phi_{i,meas}-\phi_{i,tar}$, averaging over all three phases, to be $9^{\circ}$, while the mean is only $3^{\circ}$. We estimate $3^{\circ}$ of the standard deviation is from Poisson statistical fluctuations and alignment drift in the setup over time. 

To further assess errors, we repeated the measurement 8 times for every combination of $\phi_1$, $\phi_2$, and $\phi_{3}$ at roughly $90^{\circ}$ intervals between $[0,2\pi)$; allowing us to plot a grid of the average phase measured at each point (see Appendix \ref{tomo}, Fig. \ref{90expmeas}). From these data, we calculate the mean and standard deviation of $\Delta\phi$ to be $4^{\circ}$ and $10^{\circ}$, respectively. Additionally, the average fidelity of the measured state with each target over the entire grid, including all 4 of Alice's measurement outcomes, is $F=0.93\pm0.03$. We identified some of the causes of infidelity in our system to be imperfect phase setting, imperfect phase stabilization, different measurement efficiency for the different tomography measurements, and non-equal magnitudes of the terms in the superposition; see Appendix \ref{repdata} for quantitative estimates of these effects.

In order to further assess the resolving power of our system to distinguish states with nearby phase values, for each phase ($\phi_1$, $\phi_2$, and $\phi_{3}$), we created two distributions (two liquid crystal settings) of 10 samples each, corresponding to phases differing by $7^{\circ}$ on average; we then applied a two-sample Kolmogorov-Smirnov (KS) test \cite{KStest} to test the null hypothesis (once for each phase) that all 20 samples were from the same distribution (liquid crystal setting), concluding that we can reject the null hypothesis that the data are drawn from a single distribution with $\alpha=0.05$, in other words, with a 5$\%$ probability of wrongly rejecting the null hypothesis; see Appendix \ref{KStestsec} for more information. Thus, we can estimate the total number of resolvable teleported quantum states with our system to be $(\frac{360^{\circ}}{7^{\circ}})^3\approx 136,000$.

\subsection{Doppler Shift Compensation}
\begin{figure}
\centerline{\subfloat[]{{\includegraphics[height=2.7in,width=3.5in]{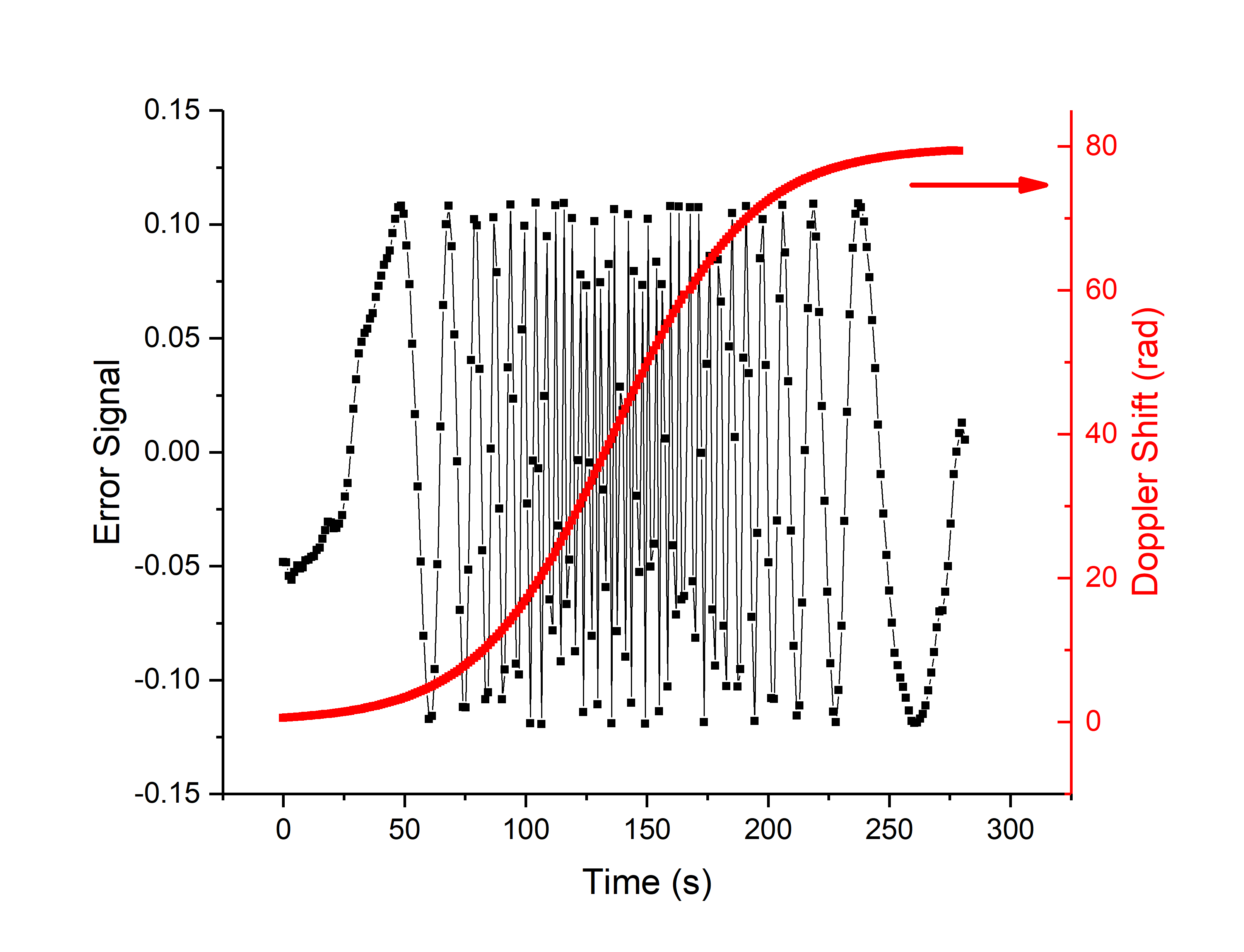}}}\subfloat[]{{\includegraphics[height=2.7in,width=3.5in]{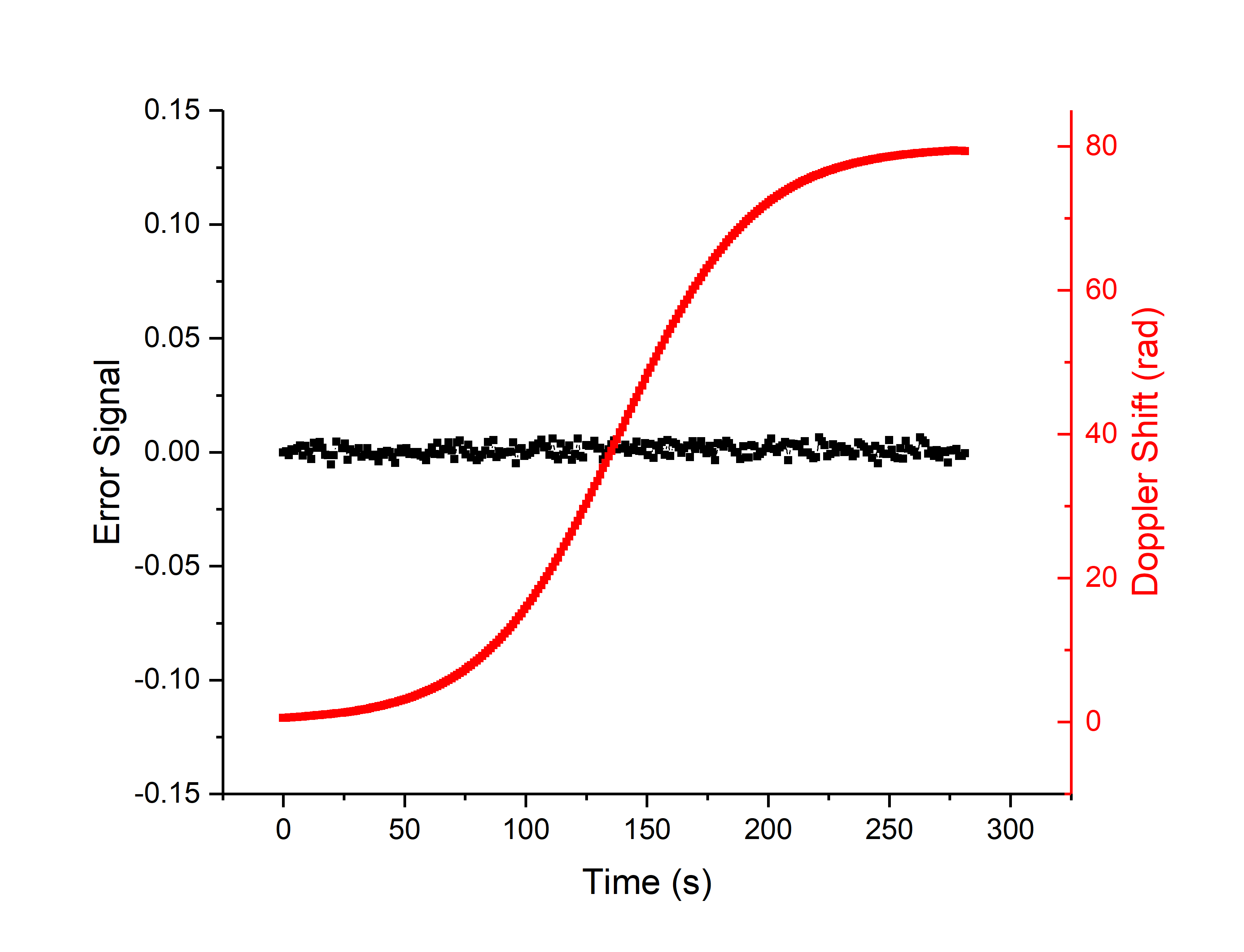}}}}\caption{Classical Doppler shift stabilization: (a) With phase stabilization off, the error signal sweeps through many interferometric fringes. (b) With phase stabilization on, the error signal is nearly constant, corresponding to $\Delta\phi=\pm1.3^{\circ}$.}\label{532doppler}
\end{figure}

Because a low-earth orbit (LEO) satellite travels at $V_r \sim 8$ km/s, the source will have moved non-negligibly between the times when the early and late time-bins  are transmitted. As the satellite approaches (recedes) this shortens (lengthens) the interval between emitted time bins from the Earth's reference frame. Uncorrected, the corresponding variation in phase (between the first two and last two terms in the state from Eqn. \ref{psibcht}) would completely obscure the phases Charles is attempting to teleport to Bob: a variation of about 80 radians is expected, depending on the time-bin separation and the orbit elevation angle. See Appendix \ref{Dopplershiftsec} for more information. To keep this Doppler shift (and any other time-varying phase shifts) from adversely affecting the protocol's performance, we developed a phase compensation system that uses a classical laser beam and proportional-integral feedback \cite{PID} to stabilize the path-length difference of the interferometers. Figure \ref{532doppler} shows the performance of the classical stabilization system while a continuously-varying, lab-simulated Doppler shift, matching that expected in a typical satellite orbit, was imposed. The standard deviation of the phase with the stabilization active is $1.3^{\circ}$.

We measured 9 tomographies while executing SDT for the same choice of $\phi_1$, $\phi_2$, and $\phi_{3}$. Without phase stabilization, we obtained an average fidelity $F=0.53\pm0.06$; with the phase stabilization turned on, we obtained an average fidelity $F=0.92\pm0.02$ and $\Delta\phi$ had a standard deviation of $14^{\circ}$. We suspect the cause of the increased phase variation during the Doppler shift to be relative drift between the pathlengths of Alice and Bob's interferometers while the Doppler shift is taking place. This should not occur during an actual implementation in space, because a Doppler shift would not occur on the time bins sent from the pump to Alice/Charles' interferometer (which is on the same platform as the source); only those sent to Bob's interferometer would experience a Doppler shift.
\subsection{Link Analysis}
\begin{figure}
\centerline{\subfloat[]{{\includegraphics[scale=0.32]{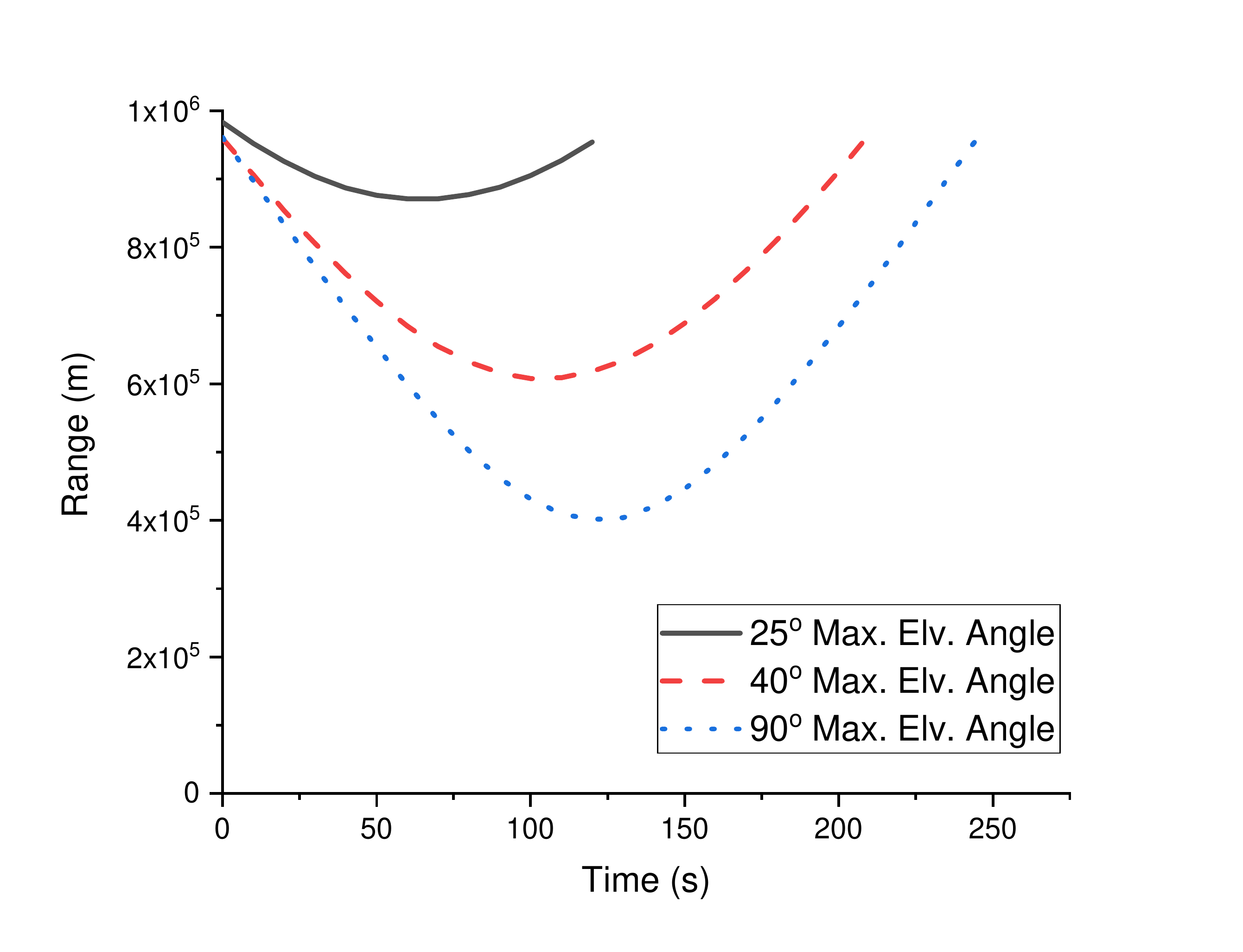}}}\subfloat[]{{\includegraphics[scale=0.32]{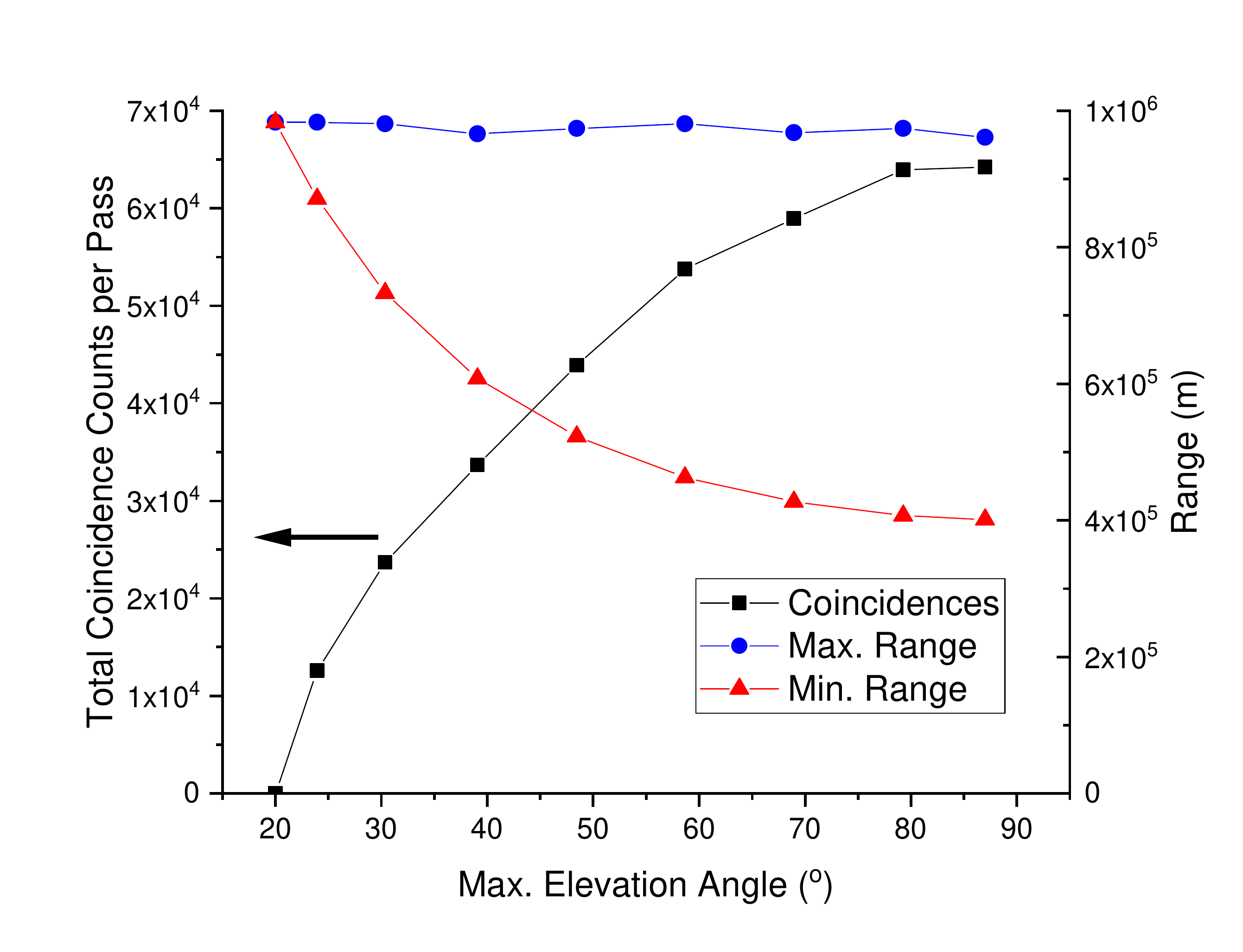}}}}\caption{Link analysis calculations: (a) Range of the satellite to a ground station at $39^{\circ}$ N latitude with a $\sim25^{\circ}$ (solid black line),  $\sim40^{\circ}$ (dashed red line), and $\sim90^{\circ}$ (dotted blue line) maximum elevation angle during the pass. (b)  On the left vertical axis, we plot the estimated total number of collected coincidence counts per orbital pass vs the maximum elevation angle of that pass (see text for underlying assumptions); on the right vertical axis, we plot the minimum and maximum range of the satellite per orbital pass.}\label{chcalc}
\end{figure}

%As shown in  Fig. \ref{chcalc}a, the elevation-angle of the satellite with respect to some ground terminal changes as it passes overhead, leading to a change in the separation between the satellite and the terminal---the ``range'';
The maximum elevation angle of a LEO satellite with respect to a ground station on earth varies from pass to pass, and the instantaneous elevation angle (defined as the angle between the horizon and the satellite) changes as the satellite passes overhead, leading to a change in the separation between the satellite and ground terminal --- the “range” (Fig. \ref{chcalc}a). With that in mind, displayed in Fig. \ref{chcalc}b, we calculate the estimated total coincidence counts per pass, maximum range per pass, and minimum range per pass versus maximum elevation angle per pass, assuming the minimum acceptable elevation angle during a pass is $20^{\circ}$ (which fixes the maximum range to around $10^6$ m, as seen in Fig. \ref{chcalc}b). For these calculations, we used simulated orbit data for all orbital parameters---the simulated satellite orbit had a 400-km altitude and $51^{\circ}$ inclination (appropriate, e.g., for the ISS), and the range as a function of time was calculated from the satellite to a ground station located at $39^{\circ}$ N latitude; see Fig. \ref{chcalc}a for example data. The Friis equation ($\eta(r)=(\pi{D_T}D_R/(4\lambda{r}))^2$) to estimate channel transmission $\eta$ as a function of range $r$  \cite{Friis,OCRD} was numerically integrated over the whole pass (for transmitting telescope diameter $D_T=0.1$ m, receiving telescope diameter $D_R=1$ m, and wavelength $\lambda=1550$ nm, with the added assumptions of a 6-dB loss for combined receiver telescope  \cite{biswas2014optical} and adaptive optics system single-mode fiber collection efficiency  \cite{chen2015experimental}, and 4-dB loss for an estimate of the analysis/detection system transmission), and assuming a 400-MHz repetition rate pump laser, pair production probability of 0.01 per pump pulse, and a 4-dB loss from the analysis/detection system in space. This experiment would require an adaptive optics correction system so the collected light can be efficiently coupled into a single-mode fiber before entering Bob's analysis/detection system. With this requirement, any turbulence is effectively converted to a reduction in transmission.
%\begin{figure}
%\centerline{\includegraphics[scale=0.6]{rangeinsetelvang70rangvtime.pdf}}\caption{Link analysis: On the left vertical axis, we plot the estimated total number of collected coincidence counts per orbital pass vs the maximum elevation angle of that pass; on the right vertical axis, we plot the minimum and maximum range of the satellite per orbital pass. (inset) Range of the satellite to a ground station at $39^{\circ}$ N latitude with a $\sim70^{\circ}$ maximum elevation angle during the pass.}\label{chcalc}
%\end{figure}
%Moreover, the simulated orbit data suggests about 3 usable passes per night for a given ground terminal at a latitude of $39^{\circ}$ N.

\begin{figure}
\centerline{\includegraphics[scale=0.6]{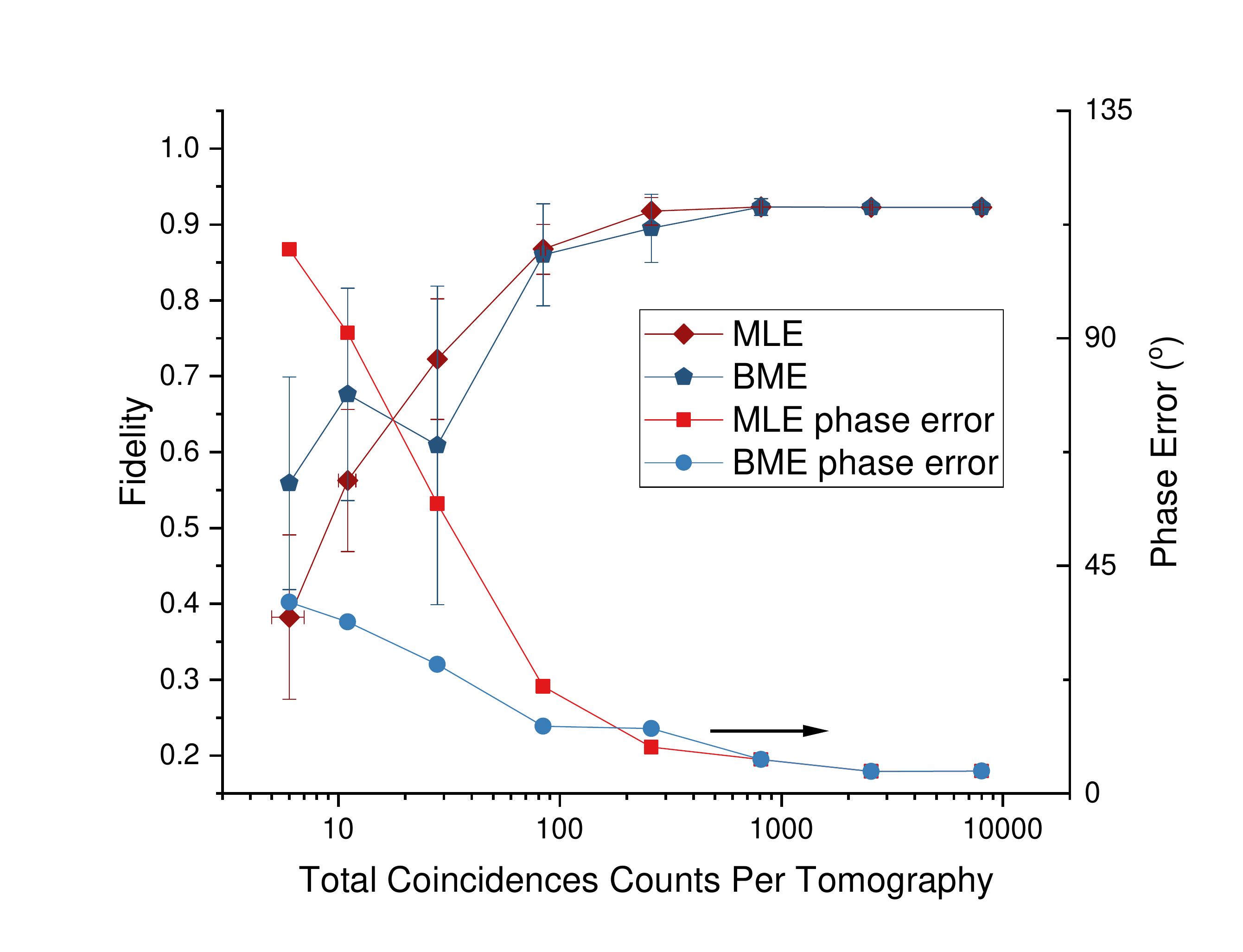}}\caption{Required counts for high fidelity SDT: The fidelity of the reconstructed state using MLE and BME, and the phase error (the standard deviation of $\Delta\phi_i\equiv\phi_{i,meas}-\phi_{i,tar}$, averaging over all three phases), as a function of the total number of coincidence counts collected per a 36-setting tomography. The fidelity and phase error was averaged over all 4 of the states produced by the different projections of Alice's detectors, $A_1$-$A_4$.}\label{BMEMLE}
\end{figure}

All tomographies analyzed thus far in the paper used maximum-likelihood estimation  (MLE) \cite{MLE}. We also analyzed some tomographic data using a Bayesian-mean-estimation (BME) approach \cite{granade2016practical}, computing the representative state as an average over all states, weighted by the likelihood that a state produced the data observed.  Using MLE yields results that are biased towards pure states \cite{ferrie2018maximum}, and this effect becomes more significant when the data used for the tomography has fewer counts, as can be seen in the low count regime in Fig. \ref{BMEMLE}. In these low-count regimes, using BME leads to results that are more reflective of the data measured. However, analyzing tomographic data using BME is more computationally intensive, especially when more than several hundred coincidence counts are collected, so MLE is often preferred with higher counts because the resultant difference between BME and MLE becomes small. 

{SDT is not inherently affected by source brightness except for slight degradation in purity from multiple pair events} \cite{chapphowest2018};{ however, to verify that the protocol is operating successfully, one needs to take a tomography of the received photons contingent on which state Alice measured ($A_1$-$A_4$); such conditional tomography is required because we are not yet implementing the final active feedforward step of the full SDT protocol. We know from previous analysis}  \cite{chapphowest2018} { that only $\sim300$ coincidence counts total per the 36 tomography measurements are needed for a reliable reconstruction (state Fidelity $>0.9$) using MLE}. {From our calculations above, all passes above 25$^{\circ}$ maximum elevation angle should produce more than 10,000 total coincidence counts per pass} (Fig. \ref{chcalc}). {Therefore, under those assumptions, a future implementation of this system in space should produce and measure more than enough coincidences to verify an implementation of SDT in a single pass. Furthermore, with currently available technology, such as adaptive optics on receiver telescopes for single-mode fiber coupling, active polarization compensation to ensure Bob and Charles have a common basis, and time-bin phase stabilization as demonstrated in this work, SDT should be implementable in a space-to-earth channel without degradation compared to our laboratory implementation.}
\section{Conclusions}
We have shown a systematic characterization of our system to execute SDT, and characterize the full volume of accessible quantum states by measuring the fidelity of states at regular intervals of phase. The phase error was measured from this characterization, along with the distinguishability of closely spaced phases. We also demonstrated the ability to operate during a Doppler shift by employing an active feedback system. Lastly, we calculate the expected coincidence counts for a range of satellite orbital passes and show that for nearly all of them we should have ample counts to reconstruct the received state faithfully.

The value in quantum communication occurs when two remote parties can coordinate to achieve some desirable task beyond the capabilities of classical communication.  Because SDT transmits only a restricted space of states, one might worry that the protocol would be insufficiently versatile to enable interesting or useful quantum processing tasks.  However, the equimodular states of SDT enable high-dimensional entanglement-based quantum cryptography \cite{chapman2019hyperentangled}; moreover, they are just the type required for quantum fingerprinting  \cite{qfp} and for blind quantum computing  \cite{BQC}. Therefore, a space-to-earth implementation of SDT would be an enabling demonstration along the path toward a useful global quantum network.

%\subsection{Data and code availability}
%The data which supports the plots of the paper and the conclusions therein, along with the computer code used to analyze the data, are available from the corresponding author upon reasonable request.

\section{Acknowledgements}
The authors acknowledge Alexander Hill for the suggestion to use a 2-sample KS test. Thanks to MIT-Lincoln Laboratory for the orbital simulation calculations. This work was primarily supported by NASA Grant No. NNX13AP35A  and NASA Grant No. NNX16AM26G. This work was also supported by a DoD, Office of Naval Research, National Defense Science and Engineering Graduate Fellowship (NDSEG).
\section{Author Contributions}
All authors contributed to experiment design and commented on manuscript. H.B. conceptualized SDT protocol. T.G. constructed the initial optical system and wrote the preliminary version of the tomography analysis code. C.K.Z. started upgrade of detection system, implemented Bayesian analysis, and calculated the tomography settings and the contributions to the loss of fidelity. J.C.C. upgraded the optical system and finished upgrade of detection system, and carried out all experiments and MLE data analysis.  J.C.C., C.K.Z., and P.G.K. wrote the manuscript.

%\section{Competing Interests}
%The authors declare no competing interests.

\appendix
\section{State Generation and Detection} \label{stategen}
To create the photons entangled in polarization and time bin, an 80-MHz mode-locked 532-nm laser (frequency doubled from 1064 nm, Spectra Physics Vanguard 2.5W 355 laser) with a pulse width $\sim$7 ps  was sent through a $\sim$2.4-ns delay to split every pump pulse into an early and late pulse, each of which coherently pumps the polarization entanglement source \cite{TBQubit}, a polarizing Sagnac interferometer with a Fresnel rhomb (used as a broadband half-wave plate), type-0 periodically poled (poling period is 7.5 $\mu$m) lithium niobate crystal, and a calcite crystal (to compensate for dispersion); the horizontal (vertical) component of the diagonally polarized pump travels (counter)clockwise through the Sagnac. Neglecting time-bins, traversing the two paths of the interferometer corresponds to this transformation \cite{Sagnac1,Sagnac2}:
\begin{equation}
\frac{\big(\ket{H}_{532}+\ket{V}_{532}\big)}{\sqrt{2}}\implies\frac{\big(\ket{V}_{810}\ket{V}_{1550}+\ket{H}_{810}\ket{H}_{1550}\big)}{\sqrt{2}}\text{,}
\end{equation}
where the subscripts are nominal central wavelengths of the photons. Sending a superposition of time bins into the polarizing Sagnac results in the state Eqn. \ref{psibcht} of the main text. 

The 532-nm pump bandwidth is 64 GHz. The downconversion bandwidth was measured by stimulated downconversion (difference-frequency generation) between a tunable 1550-nm laser and the pump \cite{DFGJSI}. The tunable 1550-nm laser was swept and a peak in the collected 810-nm counts was recorded. The peak was centered at 1551 nm (corresponding to 809.7 nm for the conjugate photons), with a full-width at half-maximum width of 1.5 nm (0.4 nm) \cite{Trentthesis}.

Due to birefringence, $\ket{H}$ and $\ket{V}$ do not exit the Sagnac source at exactly the same time. To compensate for this we inserted 0.5-mm of a-cut calcite into the 1550-nm beam path. This increased the visibility in the diagonal polarization basis from 91\% to 98\%. 

For Charles to encode his desired phases, he used $LC_A$, $LC_B$, and $LC_C$, with their fast axes located along the horizontal, horizontal, and vertical axes, respectively, to allow arbitrary phase selection over the range $[0,2\pi)$.

The 810-nm photons were detected by 4 avalanche photodiodes (Excelitas SPCM-AQ4C) with efficiency $\sim$45\%. The 1550-nm photons were detected by 4 1550-nm-optimized WSi superconducting nanowire detectors from NASA's Jet Propulsion Laboratory, with efficiency $\sim$80\% \cite{chapCLEO2017}; Bob's detector B2 had an efficiency of $\sim$40\% due to coupling fiber misalignment after installation. The outputs of the detectors were collected by a timetagger with 156-ps resolution (UQDevices UQD-Logic-16). {The symmetric heralding efficiency into single-mode fiber was $\sim$ 0.01, when including the above detection efficiency, analysis/detection system transmission ($\sim$ 0.3), and entangled-photon-source collection efficiency into single-mode fiber ($\sim$ 0.13).}

\section{Time-bin Phase Stabilization} \label{stab}
Due to environmental disturbances, temperature fluctuations, and the simulated Doppler shift, it was necessary to implement an active phase-stabilization system to simultaneously stabilize the phases between $\ket{t_1}$ and $\ket{t_2}$ in both Alice/Charles' and Bob's analyzer interferometers, relative to the pump interferometer. We directed some of the pump light, exiting the unused port of the pump delay interferometer, into the analyzer interferometers (see Fig. \ref{sdtsetup}). The pump light was vertically displaced from the 810-nm photons so it would not propagate through the liquid crystals and receive a phase shift. The light was detected by $D_1$ and $D_2$, low-bandwidth, amplified Si photodiodes (Thorlabs PDA36A), at both output ports of each interferometer. An error signal was calculated from the photodiodes:
\begin{equation}
E\equiv\frac{(I_{D_1}-\gamma{I_{D_2}})}{(I_{D_1}+\gamma{I_{D_2}})}\text{, with }\gamma\sim0.6\text{.}
\end{equation}
The factor $\gamma$ is necessary to balance the different visibilities measured in each output port, since the optics used in the analyzer are designed for the downconversion wavelengths and not the stabilization wavelength. For each analyzer interferometer, this error signal was input to a Proportional-Integral (PI) feedback algorithm with a set-point of zero and an output rate of 100 Hz. The PI algorithm output was fed to a driver to actuate a piezo-electric crystal on the translation stage of the right-angle prism inside the corresponding analyzer interferometer.

\section{Time-Bin Filtering} \label{fastand}
None of the the detectors used in this experiment were gated internally, allowing photon detection at any time. Initially, this presented a problem because there are three pulses emitted from Alice's and Bob's analyzer interferometers. For this experiment, it was necessary to implement a circuit to filter out events from the outer two pulses, because only the middle pulse contained events with a superposition of time bins.  Each pulse emitted from the interferometer has a fixed delay with respect to the input pulse, so employing an AND gate between each detector and the laser clock (with an adjustable delay) created a time filter with a width of $\sim$1 ns centered around the middle pulse \cite{chapman2019hyperentangled}.
\section{Tomographic Reconstruction} \label{tomo}
To reconstruct the state of the photons received by Bob, 36 different measurements were made by rotating the waveplates, moving the removable polarizer (some settings required a certain polarizer; see Table \ref{tomosetts}), and recording the coincidences between Alice and Bob's detectors. 

\begin{figure}
\centerline{\includegraphics[height=3in,width=2in]{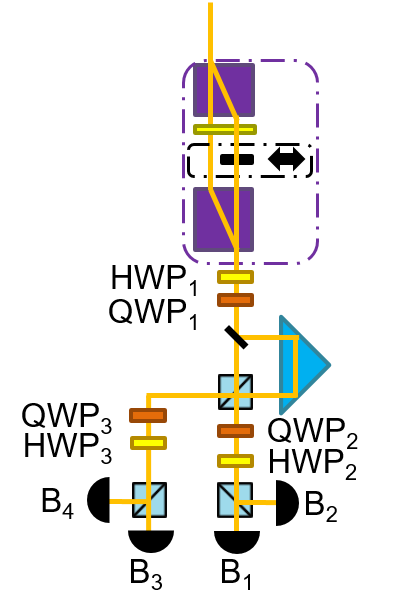}}\caption{Tomography system: This schematic shows in more detail the tomography system used to measure Bob's photon.}\label{bobtomo}
\end{figure}

The measurements performed using the setup in Fig. \ref{bobtomo} were: 
\begin{align}
&\{\ket{H},\ket{V},\ket{D},\ket{A},\ket{R},\ket{L}\}\otimes\{\ket{t_1},\ket{t_2}\}\nonumber\\
&\{\ket{H},\ket{V}\}\otimes\{\frac{1}{\sqrt{2}}(\ket{t_1}\pm i \ket{t_2}),\frac{1}{\sqrt{2}}(\ket{t_1}\pm \ket{t_2})\}\nonumber\\
&\{\frac{1}{\sqrt{2}}(\ket{Dt_1}\pm i \ket{At_2}),\frac{1}{\sqrt{2}}(\ket{Dt_1}\pm \ket{At_2}),\frac{1}{\sqrt{2}}(\ket{At_1}\pm i \ket{Dt_2}),\frac{1}{\sqrt{2}}(\ket{At_1}\pm \ket{Dt_2})\}\nonumber\\
&\{\frac{1}{\sqrt{2}}(\ket{Rt_1}\pm i \ket{Lt_2}),\frac{1}{\sqrt{2}}(\ket{Rt_1}\pm \ket{Lt_2}),\frac{1}{\sqrt{2}}(\ket{Lt_1}\pm i \ket{Rt_2}),\frac{1}{\sqrt{2}}(\ket{Lt_1}\pm \ket{Rt_2})\}\label{tomomeaseqns}
\end{align}
where D(A) is (anti-)diagonal polarization and R(L) is right(left) circular polarization. These measurements form an informationally overcomplete set in the space of interest; after data collection, they were analyzed to produce 4 density matrices (1 for each tomography conditional on which of Alice's detectors fired) using maximum-likelihood estimation \cite{MLE}.

To measure states in the first group of measurements in Eqn. \ref{tomomeaseqns}, $HWP_2$ and $HWP_3$ are rotated to 0$^{\circ}$ or 45$^{\circ}$ to project the detectors onto one time bin or the other but not superpositions of them. $HWP_1$ and $QWP_1$ in front of the interferometer are used to change what basis the PBS in the interferometer projects on to. To measure states in the second group, the beam block in the removable polarizer moves to block the orthogonal polarization. The polarization is rotated into the D/A basis and $HWP_2$ and $HWP_3$ are rotated to 22.5$^{\circ}$, so the detectors project onto superpositions of the time bins. Also, to maintain the same level of phase sensitivity across all measurements, the count time is doubled since the polarizer blocks roughly half the photons. For measurements of the third and fourth groups, $HWP_1$ and $QWP_1$ are rotated to put the PBS in the correct basis and $HWP_2$ and $HWP_3$ are rotated to 22.5$^{\circ}$ as before. To change the phase shift between $\ket{t_1}$ and $\ket{t_2}$, $QWP_2$ and $QWP_3$ are rotated. See Table  \ref{tomosetts} for exact settings.

\subsection{Tomography Measurement Efficiency Calibration} \label{effcorr}
There were 4 tomographies measured simultaneously, each conditional on one of Alice's 4 detectors. Additionally, there were 4 different simultaneous measurements because all 4 of Bob's detectors projected onto a different state during the 9 different settings for the wave plates and polarizer. To allow the use of all four of Bob's detectors for a single tomography, a measurement efficiency calibration was made periodically so the differences in the path and detection efficiencies for $B_2$-$B_4$ could be normalized to Bob's detector $B_1$. We calibrated the measurement efficiency of every measurement in the tomography with respect to detector $B_1$. This calibration consisted of five tomographies with 36 measurement settings. From these tomographies, we are able to calculate the average measurement efficiency ratio between taking the measurement with detector $B_1$ and one of the other three detectors (see the four right-most columns of Table \ref{tomostates}). Table \ref{tomostates} shows the exact mapping between the states measured in the 9- and 36-setting tomographies. A complete efficiency calibration of Bob's and Alice's measurement systems was not carried out throughout the experiment. Therefore, all tomographies measured include effects from the measurement efficiencies (from the different paths) to Bob's detector $B_1$ ($B_2$-$B_4$ were normalized to $B_1$) and the measurement efficiencies for Alice's detectors. We are able to reduce adverse effects on the measured fidelity to $\sim1\%$ by balancing the measurement efficiencies using detector alignment and by adjusting the relative probabilities for the terms in our equimodular state. Without this balancing, the fidelities would have been degraded by $\sim5\%$ to $\sim10\%$.  Effectively, the states that result from our tomography are the states collected by our detectors, not the states that enter our measurement system. Equivalent results would have been obtained for the reconstruction of states that enter our measurement system if a complete system efficiency calibration had been carried out so that the differing path efficiencies could be normalized away and the state creation elements were rebalanced accordingly.

\begin{table}
\caption{Tomography states: This table details the states that each single-photon detector is projecting onto for each setting of the tomography. Additionally, for the 9-setting tomography, it shows which coincidences are used to calibrate the relative efficiencies of Bob's 4 detectors for each tomography measurement. For example, C$_{\text{S2B1}}$ corresponds to the coincidences with Bob's detector 1 on measurement setting 2 with one of Alice's detectors.}
\centerline{\includegraphics[scale=0.5]{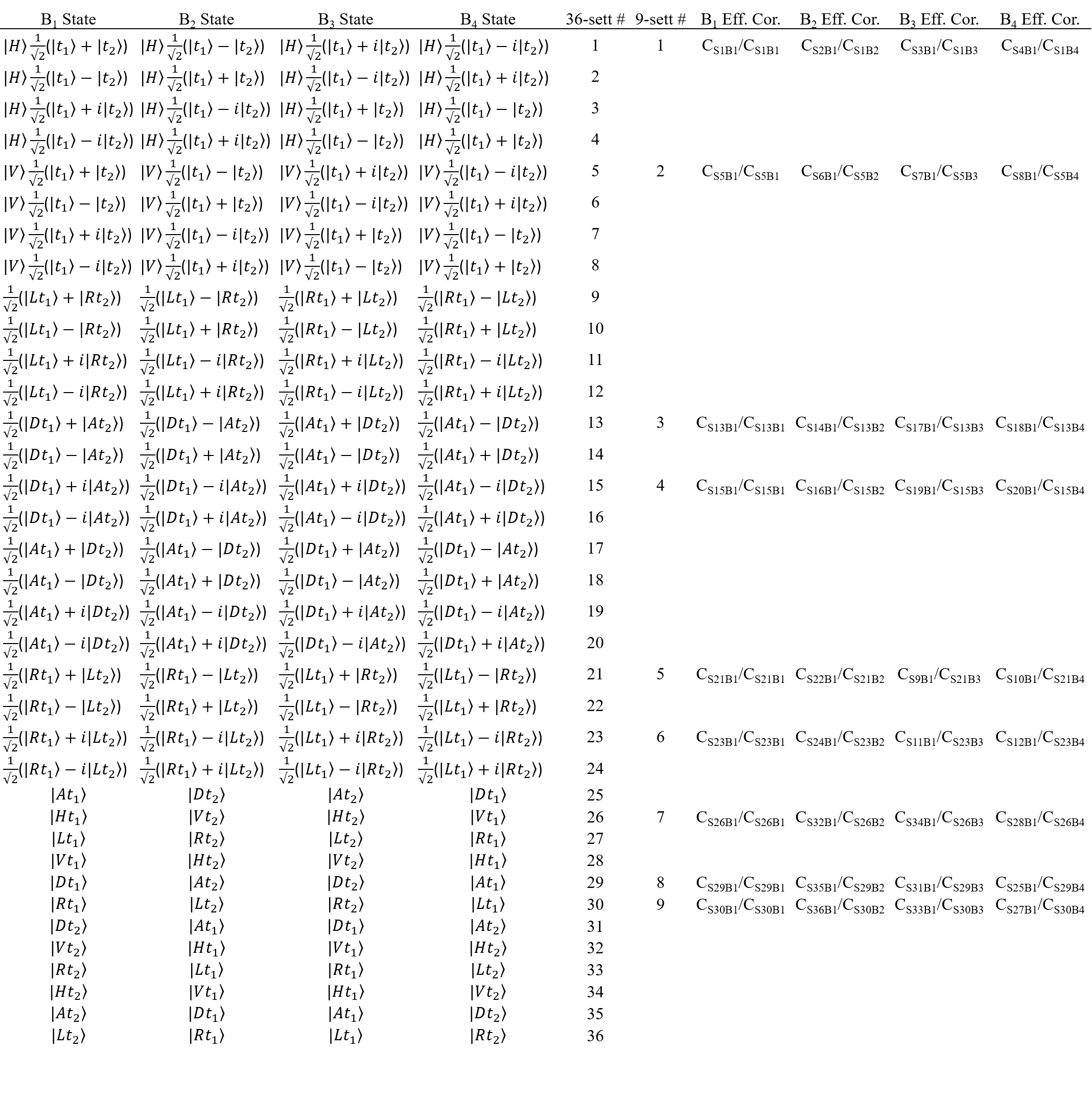}}\label{tomostates}
\end{table}
\begin{table}
\caption{Tomography settings: This table details the angles of the wave plates and positions of the polarizers for each setting of the tomography. $\alpha_1$, $\alpha_2$, $\alpha_3$ are the angles of half-wave plates 1-3; similarly $\beta_1$, $\beta_2$, $\beta_3$ are the angles of quarter-wave plates 1-3. $T_H$ and $T_V$ are transmission settings of the removable polarizer.}
\centerline{\includegraphics[scale=0.9]{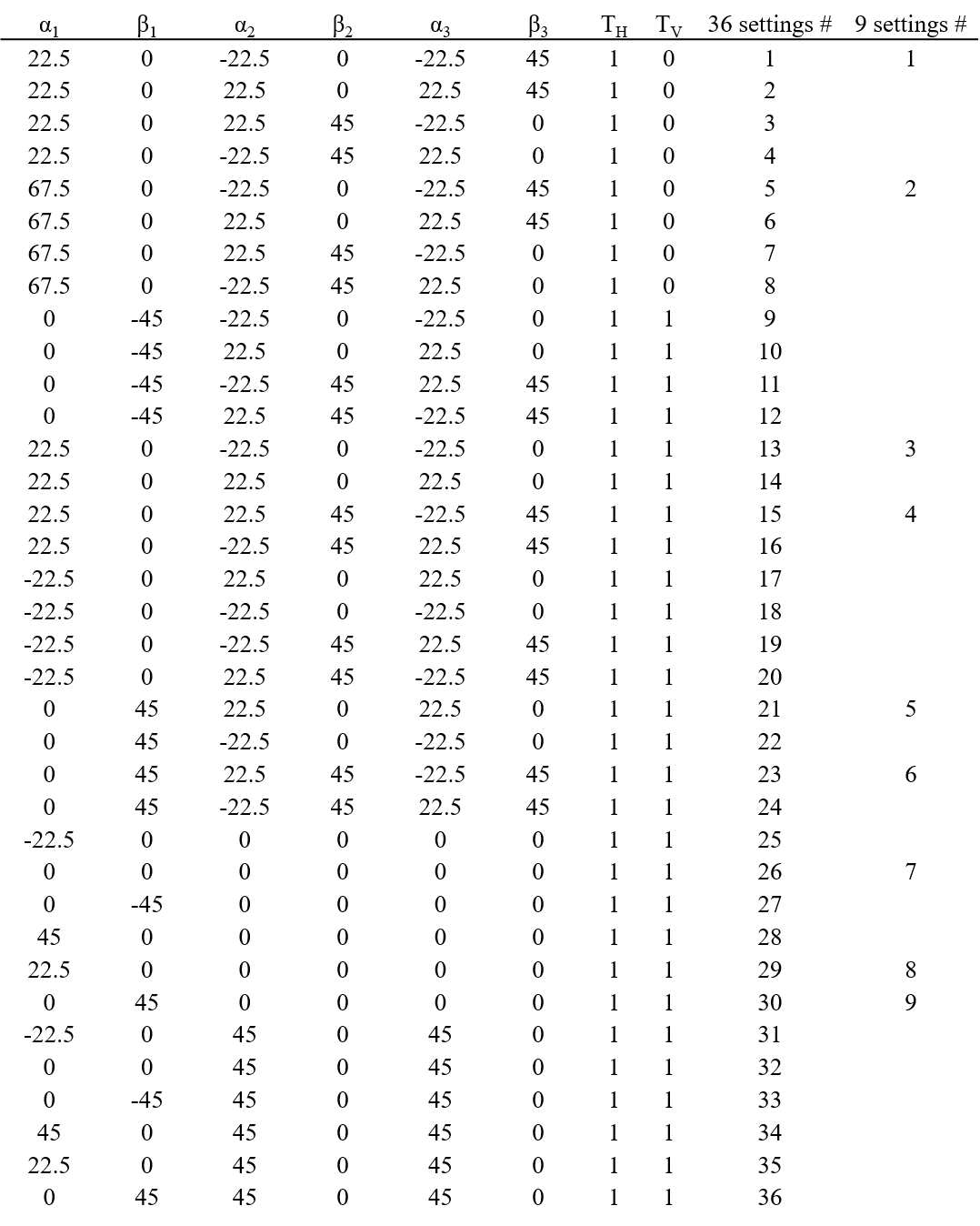}}\label{tomosetts}
\end{table}
\subsection{MLE Optical System Representation} \label{MLErep}
For maximum-likelihood state estimation, the optical system is simulated using Jones calculus and is operated on a density matrix with optimizable parameters \cite{MLE}. The density matrix, $\rho_{test}$, is constrained to represent a physical state using a Cholesky decomposition: 
\begin{equation}
\rho_{test}=U_{test}\cdot L_{test}\text{,} 
\end{equation}
where 
\begin{equation}
L_{test}\equiv
\begin{bmatrix}
t_1 &0&0&0\\
t_5+it_6&t_2&0&0\\
t_7+it_8&t_{11}+it_{12}&t_3&0\\
t_{9}+it_{10}&t_{13}+it_{14}&t_{15}+it_{16}&t_4
\end{bmatrix}
\text{and } U_{test}\equiv
\begin{bmatrix}
t_1 &t_5-it_6&t_7-it_8&t_{9}-it_{10}\\
0&t_2&t_{11}-it_{12}&t_{13}-it_{14}\\
0&0&t_3&t_{15}-it_{16}\\
0&0&0&t_4
\end{bmatrix}\text{.}
\end{equation}

We represented the optical system making the measurement as
\begin{equation}
Measp_i\equiv Setang_i\cdot{Setang_{i}}^{\dagger}
\end{equation}
  
 where
\begin{align}
 Setang_1&\equiv BPol\cdot BHWP_1 \cdot BQWP_1\cdot BHFO\cdot Interf_{P1}\cdot QWP_2\cdot HWP_2\cdot \begin{bmatrix} 1\\0\end{bmatrix}\text{,}\\
 Setang_2&\equiv BPol\cdot BHWP_1 \cdot BQWP_1\cdot BHFO\cdot Interf_{P1}\cdot QWP_2\cdot HWP_2\cdot \begin{bmatrix} 0\\1\end{bmatrix}\text{,}\\
  Setang_3&\equiv BPol\cdot BHWP_1 \cdot BQWP_1\cdot BHFO\cdot Interf_{P2}\cdot QWP_3\cdot HWP_3\cdot \begin{bmatrix} 1\\0\end{bmatrix}\text{,}\\
 Setang_4&\equiv BPol\cdot BHWP_1 \cdot BQWP_1\cdot BHFO\cdot Interf_{P2}\cdot QWP_3\cdot HWP_3\cdot \begin{bmatrix} 0\\1\end{bmatrix}\text{,}
\end{align}
\begin{equation}
Interf_{P1}\equiv
\begin{bmatrix}
1&0\\
0&0\\
0&0\\
0&1
\end{bmatrix}\text{,}
\end{equation}
\begin{equation}
Interf_{P2}\equiv
\begin{bmatrix}
0&0\\
1&0\\
0&1\\
0&0
\end{bmatrix}\text{,}
\end{equation}
\begin{equation}
BHFO\equiv
\begin{bmatrix}
1&0&0&0\\
0&-1&0&0\\
0&0&1&0\\
0&0&0&-1
\end{bmatrix}\text{,}
\end{equation}
\begin{equation}
BPol\equiv
\begin{bmatrix}
(2T_H-T_V)T_H&0&0&0\\
0&(2T_V-T_H)T_V&0&0\\
0&0&(2T_H-T_V)T_H&0\\
0&0&0&(2T_V-T_H)T_V
\end{bmatrix}\text{,}
\end{equation}
\begin{equation}
BHWP_1\equiv
\begin{bmatrix}
\cos(2\alpha_1)&-2\cos(\alpha_1)\sin(\alpha_1)&0&0\\
-2\cos(\alpha_1)\sin(\alpha_1)&-\cos2(\alpha_1)&0&0\\
0&0&\cos(2\alpha_1)&-2\cos(\alpha_1)\sin(\alpha_1)\\
0&0&-2\cos(\alpha_1)\sin(\alpha_1)&-\cos(2\alpha_1)
\end{bmatrix}\text{,}
\end{equation}
\begin{equation}
BQWP_1\equiv
\begin{bmatrix}
\cos^2(\beta_1)+i\sin^2(\beta_1)&(i-1)\cos(\beta_1)\sin(\beta_1)&0&0\\
(i-1)\cos(\beta_1)\sin(\beta_1)&i\cos^2(\beta_1)+\sin^2(\beta_1)&0&0\\
0&0&\cos^2(\beta_1)+i\sin^2(\beta_1)&(i-1)\cos(\beta_1)\sin(\beta_1)\\
0&0&(i-1)\cos(\beta_1)\sin(\beta_1)&i\cos^2(\beta_1)+\sin^2(\beta_1)
\end{bmatrix}
\end{equation}\text{,}
\begin{equation}
HWP_{j}\equiv
\begin{bmatrix}
\cos(2\alpha_{j})&-2\cos(\alpha_{j})\sin(\alpha_{j})\\
-2\cos(\alpha_{j})\sin(\alpha_{j})&-\cos(2\alpha_{j})
\end{bmatrix}
\text{for }j\equiv\{2,3\}\text{, and}
\end{equation}
\begin{equation}
QWP_{j}\equiv
\begin{bmatrix}
\cos^2(\beta_{j})+i\sin^2(\beta_{j})&(i-1)\cos(\beta_{j})\sin(\beta_{j})\\
(i-1)\cos(\beta_{j})\sin(\beta_{j})&i\cos^2(\beta_{j})+\sin^2(\beta_{j})
\end{bmatrix}
\text{for }j=\{2,3\}\text{.}
\end{equation}
$\alpha_1$, $\alpha_2$, $\alpha_3$ are the angles of half-wave plates 1-3, and $\beta_1$, $\beta_2$, $\beta_3$ are the angles of quarter-wave plates 1-3. $T_H$ and $T_V$ are the transmission settings of the removable polarizer; for example, if $T_H=1$ and $T_V=0$, the removable polarizer is positioned such that $\ket{H}$ is transmitted while $\ket{V}$ is blocked. The angles and positions of these elements during a tomography are listed in Table \ref{tomosetts}. 

\subsection{Representative Data and Error Analysis}\label{repdata}
\begin{figure}
\centerline{\includegraphics[height=4.7in,width=9in]{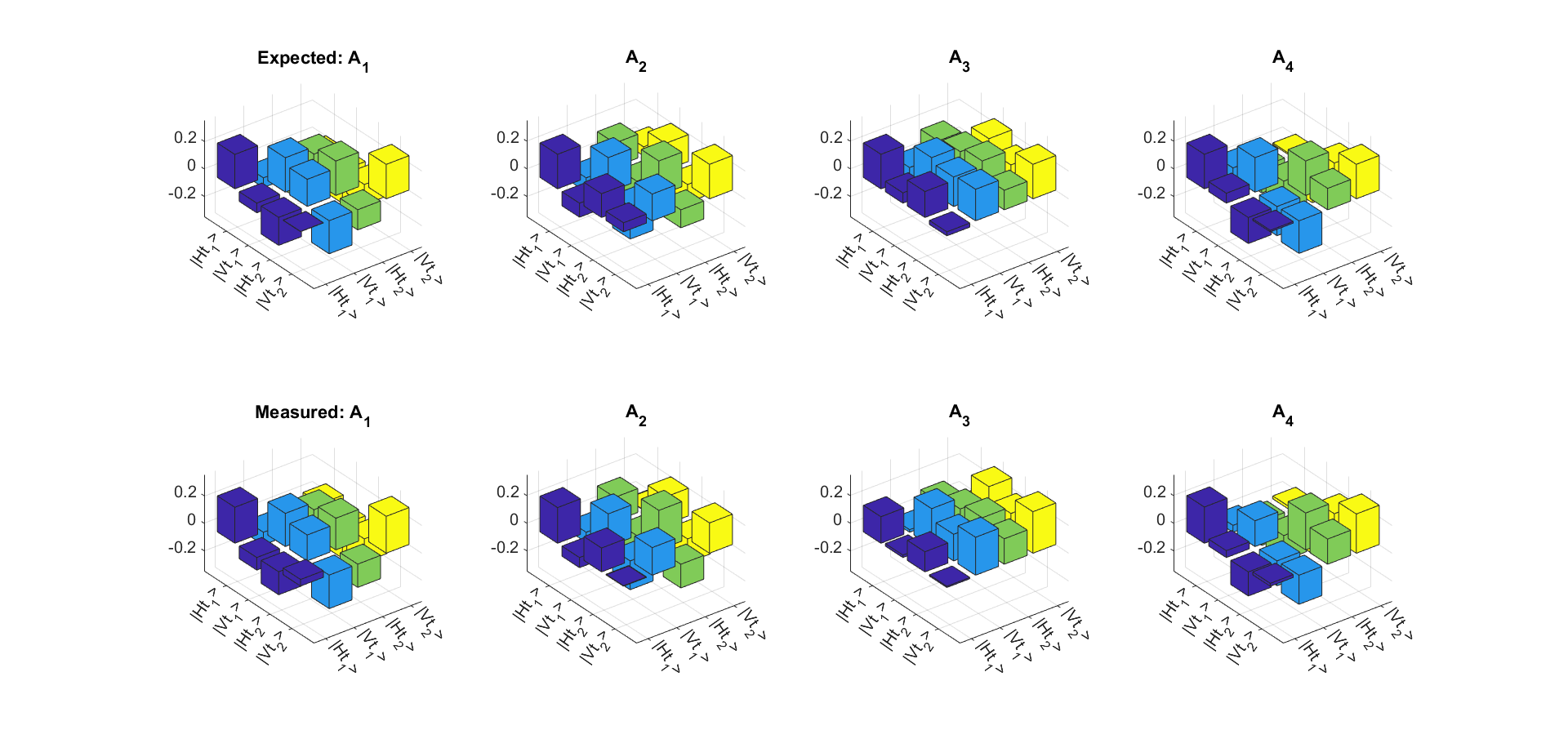}}\caption{Representative density matrix: Real part of expected and reconstructed density matrices for the states received by Bob, labeled by Alice's measurement outcome. The raw counts data is shown in Table \ref{countfile}. Fidelities between measured and expected states for $A_1$, $A_2$, $A_3$, and $A_4$ are 0.94, 0.93, 0.95, and 0.94, respectively.}\label{densmatfig}
\end{figure}

\begin{table}
\caption{Representative tomography data: The singles and coincidence counts measured for each setting in a tomography (the columns) for one of the states measured in Fig. \ref{45expmeas}, specifically the state in Fig. \ref{densmatfig}.}
\centerline{\includegraphics[scale=0.9]{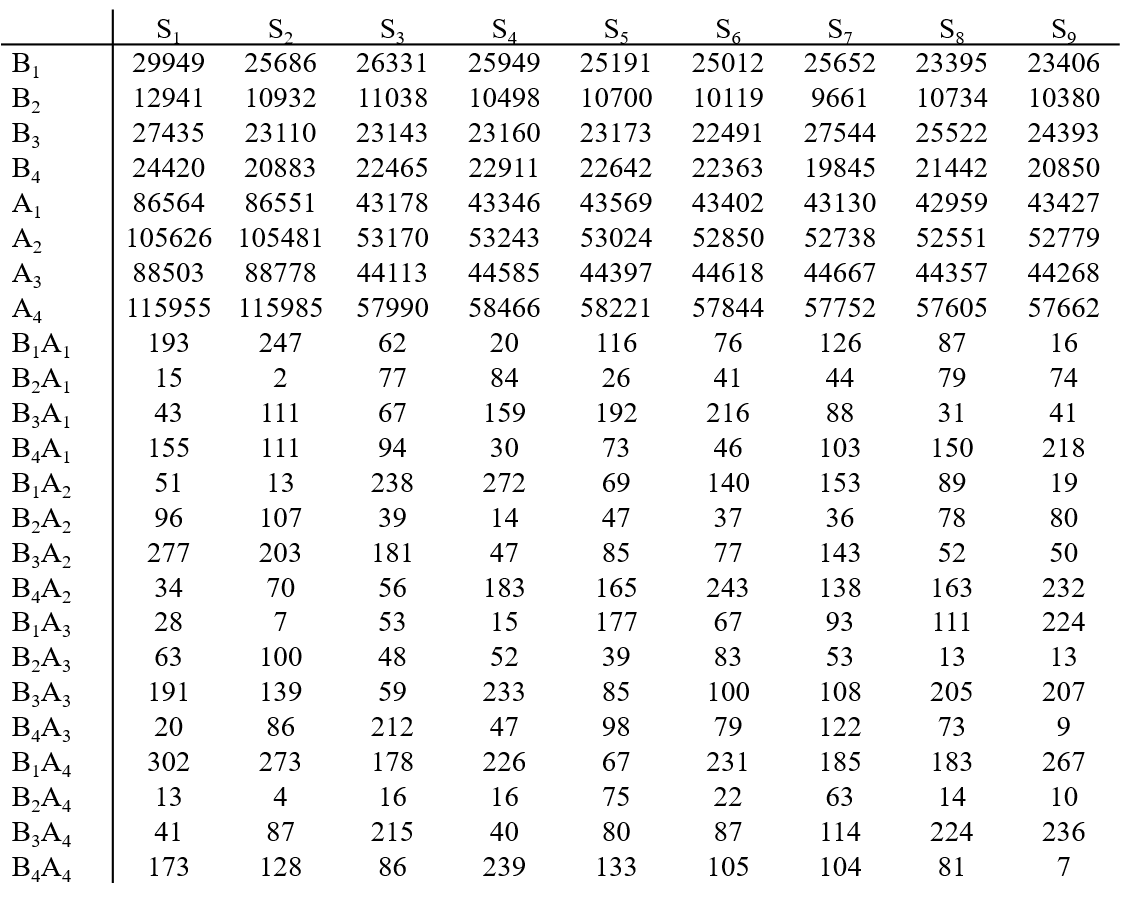}}\label{countfile}
\end{table}
The real parts of the density matrices of the expected and received states for a typical superdense teleported state are shown in Figure \ref{densmatfig}. The received states were reconstructed from the data in Table \ref{countfile}. This data was taken while counting for 15 seconds for each setting, except the settings that used the movable polarizer in front of Bob's interferometer (9-settings \# 1-2, 36-settings \# 1-8), which used a count time of 30 seconds. The pump power was $\sim$ 0.5 mW at the PPLN crystal.

From measurements of components in our system and detailed numerical simulation, we find $\sim1\%$ drop in fidelity from imperfect polarizer extinction ratios, $\sim1\%$ from imperfect LC basis and phase settings, $\sim1\%$ from unbalanced measurement efficiencies, $\sim1\%$ from imperfect time-bin qubit purity, and $\sim2\%$ from imperfect H/V and D/A visibility in the removable polarizer.

\begin{figure}
\centerline{\includegraphics[height=4.7in,width=9in]{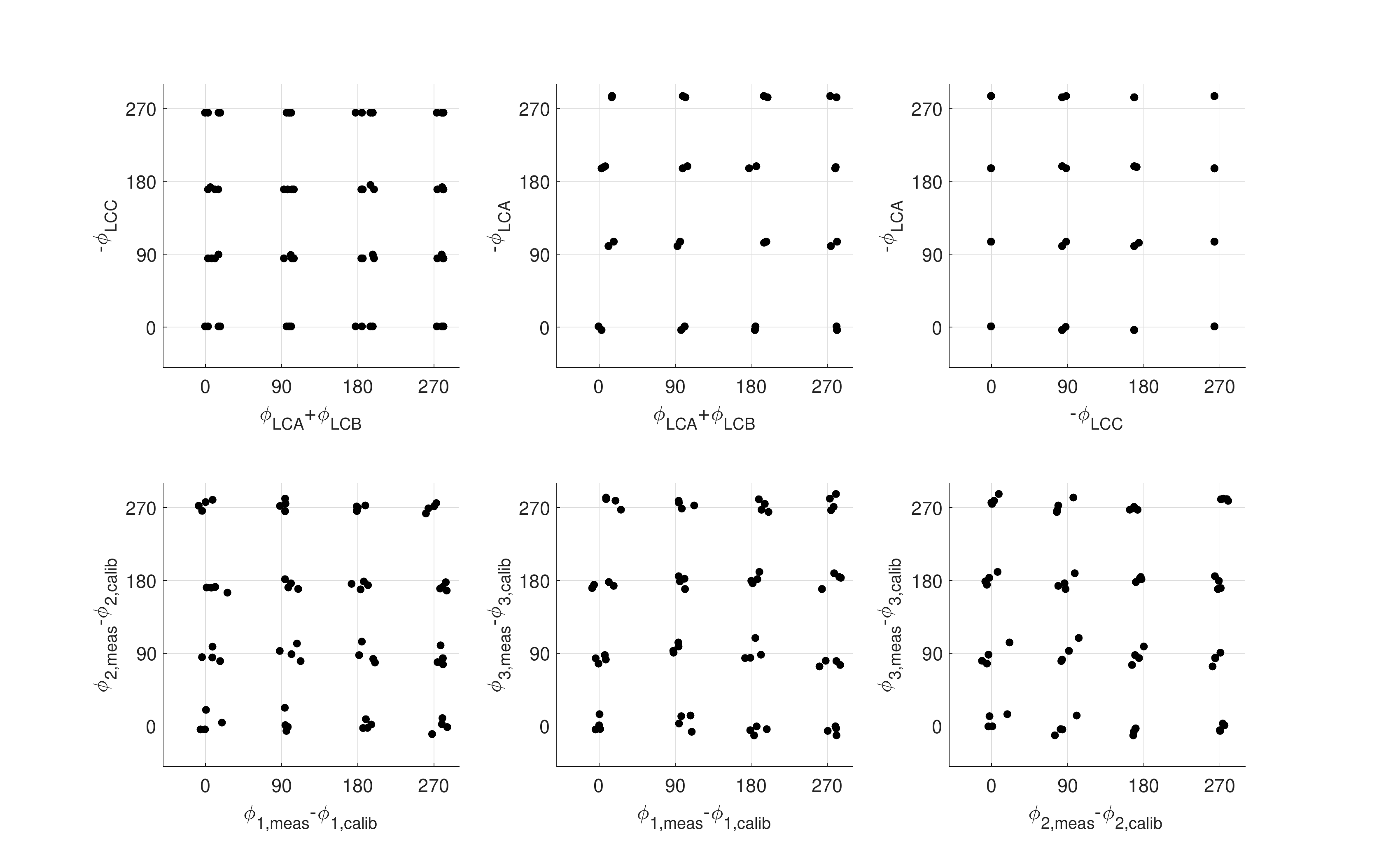}}\caption{Projections onto $90^{\circ}$ grid: Averaged projections onto $90^{\circ}$ grid over 8  independent measurements of each grid.  All axes are in units of degrees.  (top row) Target projections based on liquid crystal phase calibration. (bottom row) Measured phase projections after subtracting calibrated phase offset.}\label{90expmeas}
\end{figure}
\subsection{Full State Tomography}\label{fulltomosec}
\begin{figure}
\centerline{\includegraphics[scale=0.6]{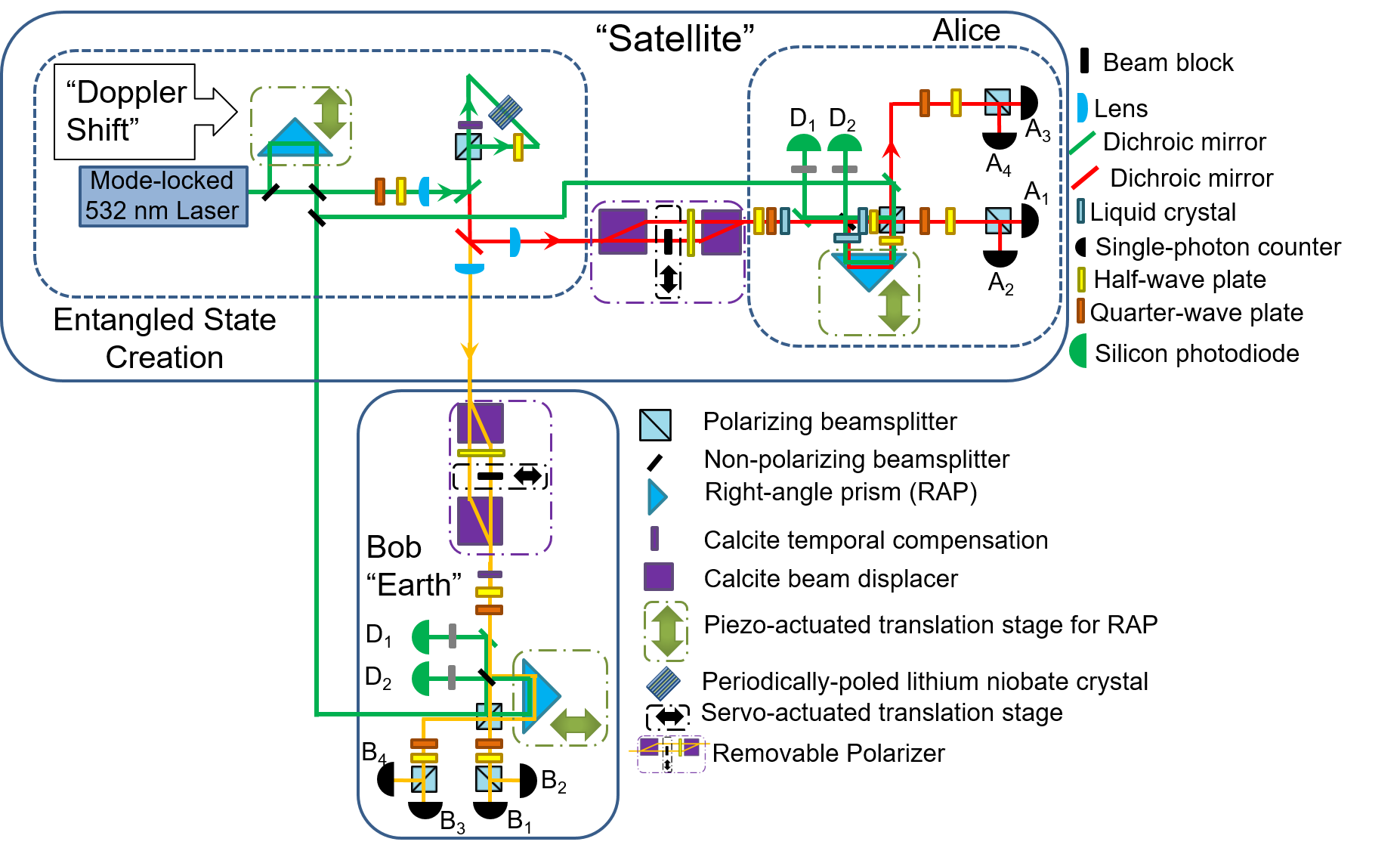}}\caption{Full state tomography optical setup: To measure the full state tomography of both photons, several half-wave plates, quarter-wave plates, and a removable polarizer were added to Alice's/Charles' measurement setup to enable the same measurements that are performed on Bob's photon.}\label{fulltomosetup}
\end{figure}

To measure the total joint state of the entangled photon pairs, a few additions were made to the system to allow a tomography to be measured on Alice/Charles' photon: a half- and a quarter-wave plate were added before Alice/Charles' interferometer and then a quarter-wave plate was added to each output port of the interferometer; additionally, a removable polarizer was added before the interferometer. The setup diagram during this measurement is in Fig. \ref{fulltomosetup}. The tomography was measured using 36 (Alice's/Charles' settings) x 36 (Bob's settings) = 1296 settings.
\section{Liquid Crystal Calibration} \label{LCcalib}
To calibrate the phase applied by each liquid crystal for each driving voltage, a tomography was measured on Bob's photon (as above) conditioned on detection of Alice's photon by detector $A_1$; the phase between H and V was then extracted from the density matrix. This is distinctly different from the phase extraction used in the analysis of the SDT protocol trials---in that case all phases were extracted, including between the time bins. Here, effectively only a polarization tomography is conducted to measure the phase between H and V applied by the liquid crystal. Additionally, to reduce phase error as much as possible, it was necessary to periodically ($\sim$ 2 days) recalibrate the phase applied by the liquid crystals as measured from the tomography system. Otherwise a drift as much as $20^{\circ}$ is observed, due to an induced change in the phase extraction from a varying measurement-efficiency imbalance. The tomography involves many projective measurements, each with a different efficiency, and such differences can modify the extracted phase values.
\section{Two-sample KS tests}\label{KStestsec}
\begin{figure}
\centerline{\includegraphics[height=4.7in,width=9in]{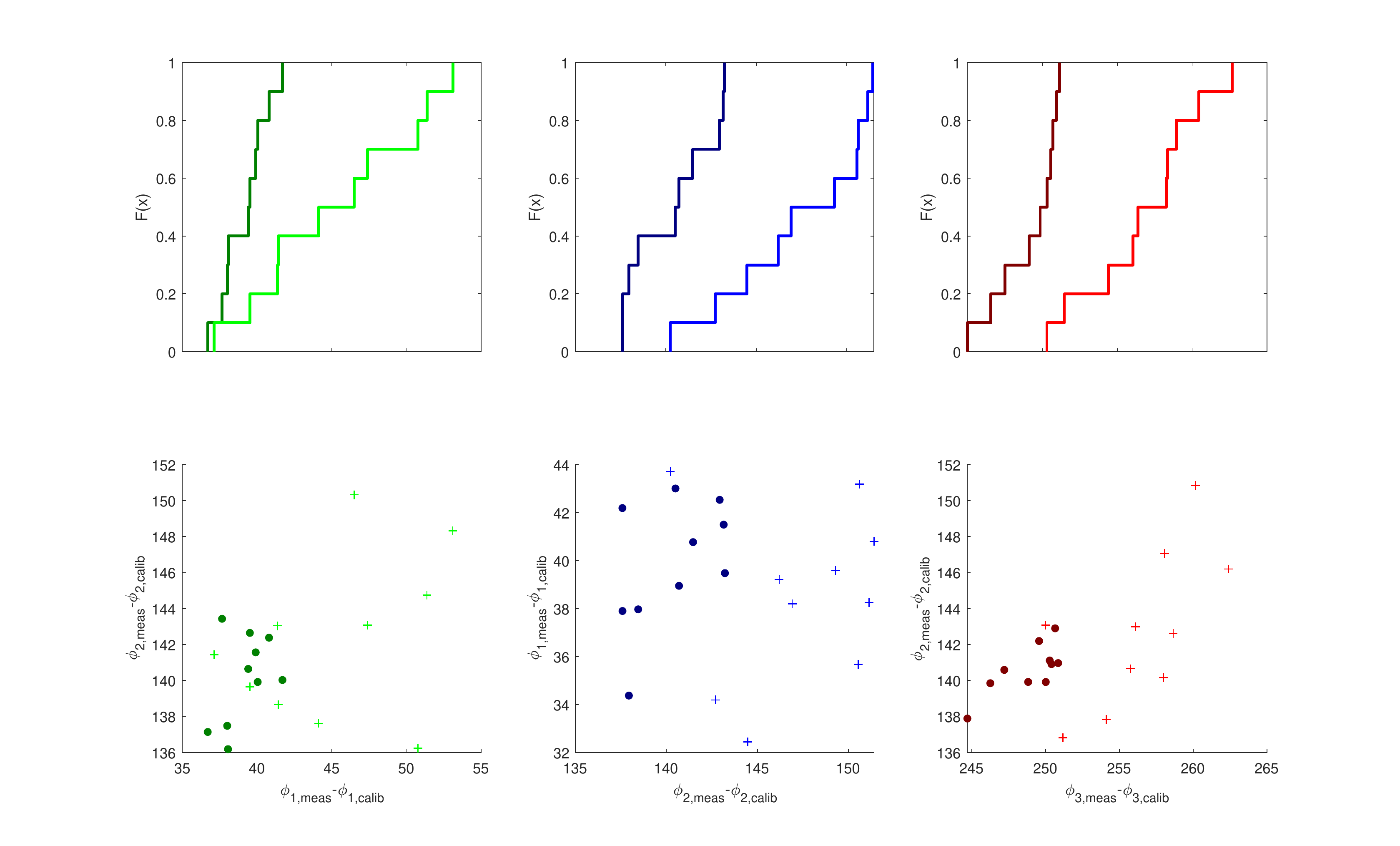}}\caption{Two-sample KS tests of  very closely spaced distributions in phase space: (top row) empirical CDF of phase space data along horizontal axis. (bottom row) Measured points in phase space. Crosses (dots) are measurements of the (un)displaced distributions. The distributions of crosses (dots) are centered over 45$^{\circ}$(39$^{\circ}$), 147$^{\circ}$(140$^{\circ}$), 256$^{\circ}$(249$^{\circ}$) for $\phi_1$, $\phi_2$, and $\phi_3$, respectively.}\label{5degKStest}
\end{figure}

\begin{figure}
\centerline{\includegraphics[height=4.7in,width=9in]{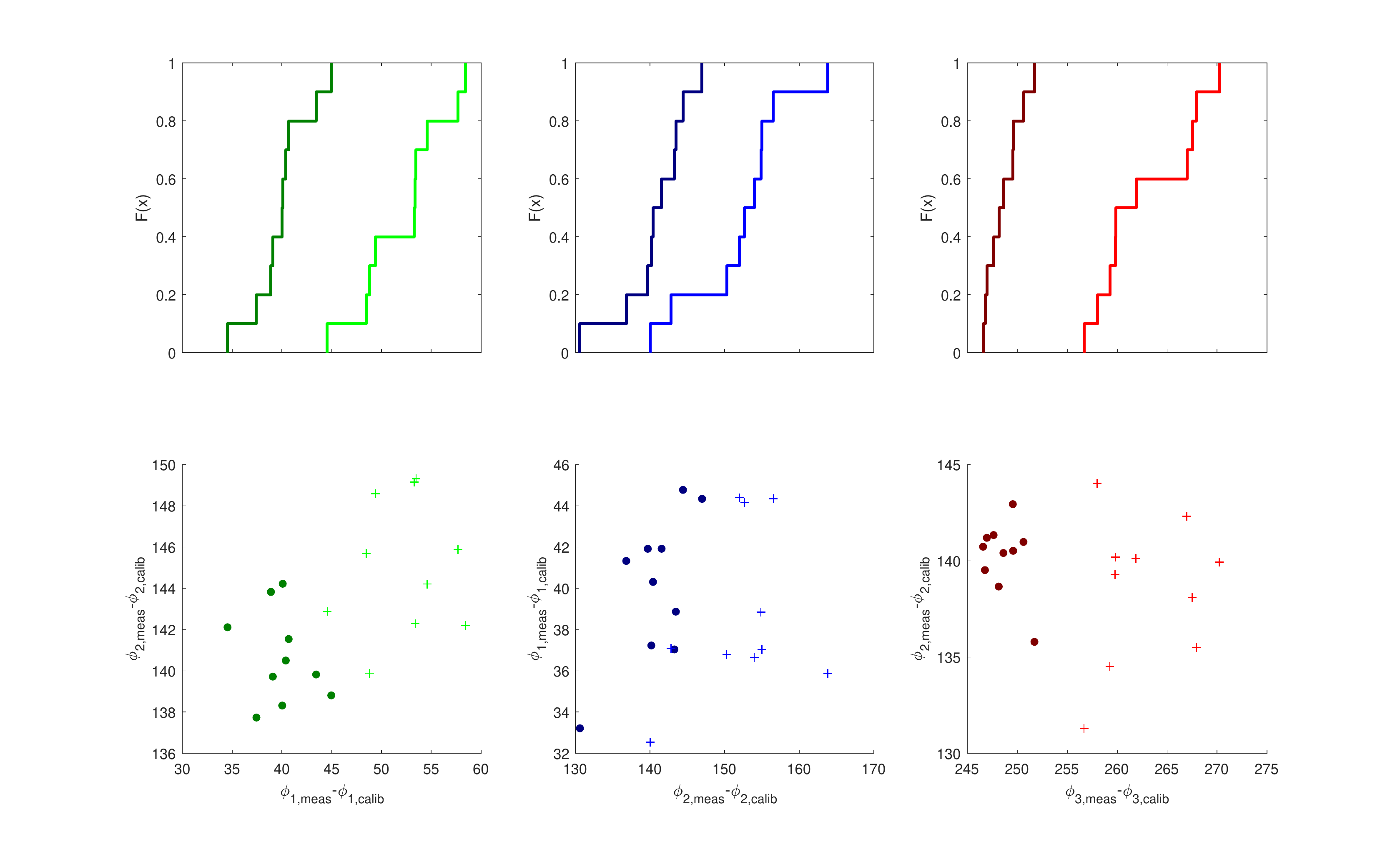}}\caption{Two-sample KS tests of moderately closely spaced distributions in phase space: (top row) empirical CDF of phase space data along horizontal axis.(bottom row) Measured points in phase space. Crosses (dots) are measurements of the (un)displaced distributions. The distributions of crosses (dots) are centered over 52$^{\circ}$(39$^{\circ}$), 152$^{\circ}$(141$^{\circ}$), 263$^{\circ}$(249$^{\circ}$) for $\phi_1$, $\phi_2$, and $\phi_3$, respectively.}\label{10degKStest}
\end{figure}

In order to assess the resolving power of our system to distinguish states with nearby phase values, for each phase ($\phi_1$, $\phi_2$, and $\phi_{3}$), we created distributions for two closely spaced phase settings (two liquid crystal settings) of 10 samples each; we then applied a two-sample Kolmogorov-Smirnov (KS) test \cite{KStest} to test the null hypothesis (once for each phase) that all 20 samples were from the same distribution (liquid crystal setting). The distributions and their empirical cumulative distribution functions (CDF) are shown in Fig. \ref{5degKStest}. These distributions had a standard deviation of $3^{\circ}$ and were, on average, separated by $7^{\circ}.$  The two-sample KS test statistic is 
\begin{equation}D_{k,l}=\sup_{x}\big| F_{1,k}(x)- F_{2,l}(x)\big|\text{,}
\end{equation} where $F_{1,k}$ and  $F_{2,l}$ are the empirical distribution functions of the first and second sample, respectively. The null hypothesis is rejected with a confidence level of $\alpha$ if
\begin{equation} D_{k,l}> c(\alpha)\sqrt{\frac{k+l}{kl}}\text{, where }c(\alpha)\equiv\sqrt{-\frac{1}{2}ln\Big(\frac{\alpha}{2}\Big)}\text{,}
\end{equation}
and $k$ and $l$ are the number of samples in each distribution. We applied the two-sample KS test to the distributions shown in Fig. \ref{5degKStest}, concluding that we can reject the null hypothesis that the data are drawn from a single distribution with $\alpha=0.05$, in other words, with a 5$\%$ probability of wrongly rejecting the null hypothesis.
We also applied the two-sample KS test to the two distributions shown in Fig. \ref{10degKStest}; these distributions had a standard deviation of $5^{\circ}$, with means separated by $13^{\circ}$. After applying the two-sample KS test, we reject the null hypothesis that they are the same distribution with $\alpha=0.005$.
\section{Doppler Shift}\label{Dopplershiftsec}
\begin{figure}
\centerline{\subfloat[]{{\includegraphics[scale=0.5]{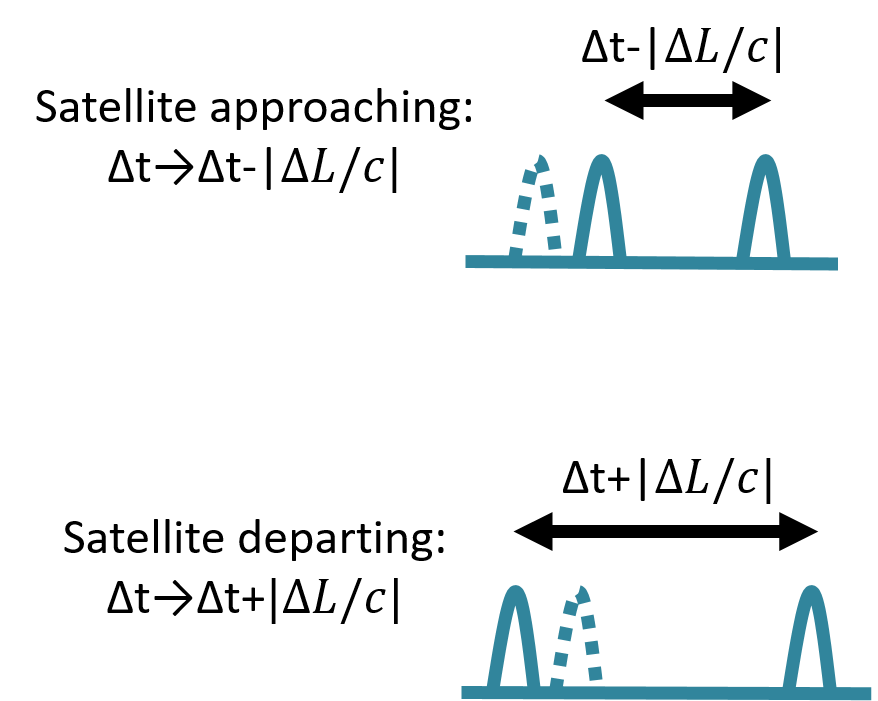}}}\subfloat[]{{\includegraphics[scale=0.4]{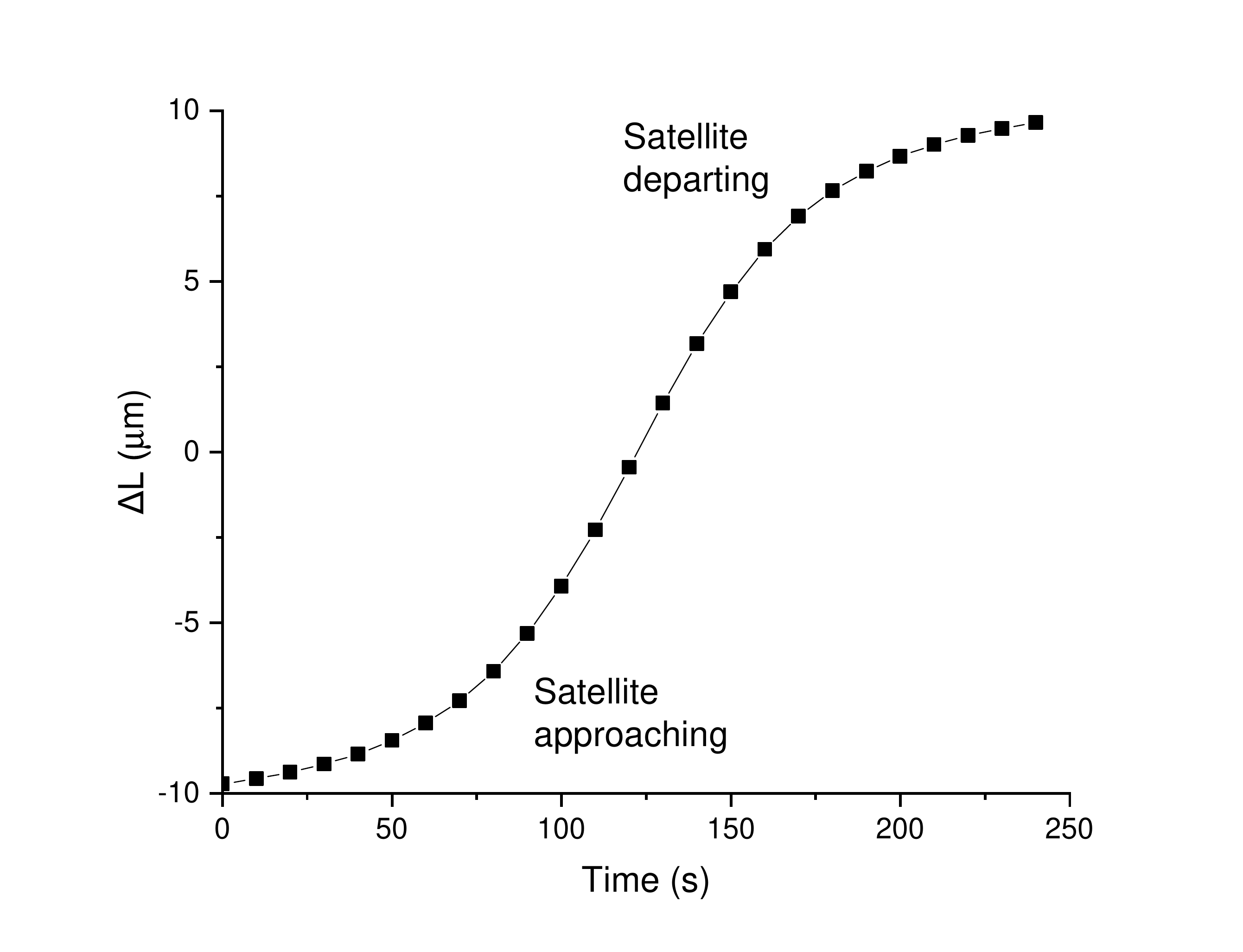}}}}\caption{Expected Doppler shift: (a) Pictorial explanation of effect of Doppler shift on time bins. (b) Expected Doppler shift for overhead orbit ($\sim90^{\circ}$ elevation angle) of satellite with velocity 7.7 km/s.}\label{Doppfigexplan}
\end{figure}

The Doppler-effect-induced phase shift is dependent on many orbital parameters, including the elevation angle of the orbit, which changes per pass and is at a maximum for passes directly overhead. Calculations using the relativistic longitudinal Doppler shift equation  \cite{reldop} show an expected shift (see Fig. \ref{Doppfigexplan}) of 
\begin{equation}
    \Delta{t}(t)=\Bigg(\sqrt{\frac{1+\frac{V_{sat}(t)}{c}}{1-\frac{V_{sat}(t)}{c}}}-1\Bigg)(1.5 \text{ ns})\text{,}\label{doppeqn}
\end{equation}
assuming time bins separated by 1.5 ns and that the maximum elevation angle during a pass for the orbit of the simulated satellite is about $90^{\circ}$. If acquisition starts and stops at a $20^{\circ}$ elevation angle, then the total $\Delta{t}$ from $t_{start}$ to $t_{stop}$ is  $\Delta{t}(t_{stop})-\Delta{t}(t_{start})=43$ fs (or $\Delta{L}=12.8\mu{m}$).

We implemented an in-lab simulation of this Doppler shift, during our compensation system testing, by moving a piezo-actuated translation stage which controlled the position of the pump's right-angle prism with a distance-vs-time profile matching Eqn. \ref{doppeqn}, as in Fig. \ref{Doppfigexplan}.
There is also a Doppler shift on the frequency on the photons; however, the frequency shift is negligible since the photon bandwidth is $\sim1$ nm and \begin{equation}\gamma\equiv\frac{1}{\sqrt{1-(\frac{V_{sat}}{c})^2}}=1.00000000033\text{,}
\end{equation}i.e., quite close to 1 for $V_{sat}=7.7$ km/s.

% Figure \ref{532doppler} shows the performance of the classical stabilization system while a lab-simulated Doppler shift, matching that expected in a typical ISS orbit, was imposed. The standard deviation of the phase with the stabilization active is $1.3^{\circ}$.
%\begin{figure}
%\centerline{\subfloat[]{{\includegraphics[height=2.7in,width=3.5in]{532Dopplerstaboff.png}}}\subfloat[]{{\includegraphics[height=2.7in,width=3.5in]{532Dopplerstabon.png}}}}\caption{Classical Doppler shift stabilization: (a) With phase stabilization off, the error signal sweeps through many interferometric fringes. (b) With phase stabilization on, the error signal is nearly constant, corresponding to $\Delta\phi=\pm1.3^{\circ}$.}\label{532doppler}
%\end{figure}
\clearpage

%\bibliography{referencesv2}{}

\end{document}